\documentclass{article}
\usepackage{amssymb}
\usepackage{amsfonts}
\usepackage{amsmath}
\usepackage{epsfig}

\newcommand{\be}{\begin{eqnarray}}
\newcommand{\ee}{\end{eqnarray}}
\newcommand{\etal}{{\it et al.}}
\def\nue{{\nu_e}}
\def\anue{{\bar\nu_e}}
\def\numu{{\nu_{\mu}}}
\def\anumu{{\bar\nu_{\mu}}}
\def\nutau{{\nu_{\tau}}}
\def\anutau{{\bar\nu_{\tau}}}
\newcommand{\ms}{\Delta m^2_{21}}
\newcommand{\ma}{\Delta m^2_{31}}

\newcommand{\sss}{\sin^2 \theta_{12}}
\newcommand{\sch}{\sin^2 \theta_{13}}

\def\ltap{\ \raisebox{-.4ex}{\rlap{$\sim$}} \raisebox{.4ex}{$<$}\ }
\def\gtap{\ \raisebox{-.4ex}{\rlap{$\sim$}} \raisebox{.4ex}{$>$}\ }
\newcommand{\sig}{$3\sigma$}
\newcommand{\chr}{\mbox{$\breve{\rm C}$erenkov~}}
\newcommand{\apee}{\bar P_{ee}}

\newcommand{\apxe}{\bar P_{xe}}

\begin{document}

\title{
\begin{flushright}
\small{hep-ph/0703092}\\
\small{HRI-P-07-03-001}\\
\small{OUTP 0703P}\\
\end{flushright}
\bigskip
Turbulent Supernova Shock Waves and the Sterile Neutrino
Signature in Megaton Water Detectors}
\author{Sandhya Choubey$^{1}$\thanks{email: \tt sandhya@mri.ernet.in}~,
N. P. Harries$^2$\thanks{email: \tt n.harries1@physics.ox.ac.uk}~,
G.G. Ross$^2$\thanks{email: \tt g.ross1@physics.ox.ac.uk}~\\\\
$^1${\normalsize \it Harish-Chandra Research Institute,} \\
{\normalsize \it Chhatnag Road, Jhunsi, Allahabad  211 019, INDIA}\\
\\
$^2${\normalsize \it The Rudolf Peierls Centre for Theoretical Physics,}\\
{\normalsize \it University of Oxford, 1 Keble Road, Oxford, OX1 3NP, UK}}

\date{}
\maketitle

\begin{abstract}

The signatures of sterile neutrinos in the supernova neutrino signal
in megaton water Cerenkov detectors are studied. Time dependent
modulation of the neutrino signal emerging from the sharp
changes in the oscillation probability due to shock waves
is shown to be a smoking gun for the existence of
sterile neutrinos. These modulations and indeed the entire
neutrino oscillation signal is found to be different
for the case with just three active neutrinos and the cases
where there are additional sterile species mixed with the
active neutrinos. The effect of turbulence is taken into account
and it is found that the effect of the shock waves, while modifed, remain significant and measurable.
Supernova neutrino signals in
water detectors can therefore give unambiguous
proof for the existence of sterile neutrinos, the sensitivity extending beyond that for terrestial neutrino
experiments. In addition the time dependent
modulations in the signal due to shock waves can be used to
trace the evolution of the shock wave inside the supernova.

\end{abstract}

\newpage

\section{Introduction}

The detection of neutrinos from SN1987A in the
Large Megallenic Cloud remains a major landmark in neutrino
physics and astrophysics. The 11 events detected in
Kamiokande \cite{k2sn1987a} and 8 events in IMB \cite{imbsn1987a}
stimulated a plethora of research papers exploring both type-II supernova dynamics
and neutrino properties \cite{reviewsn1987a}.
A supernova explosion within our own galaxy will generate
tens of thousands of events in the currently running
and proposed
neutrino detectors and hence is expected to significantly
improve our understanding of the type-II supernova explosion
mechanism on one hand and neutrino physics on the other.

Our knowledge on the pattern of neutrino mass and mixing
has seen tremendous improvement from the results of
a series of outstanding
experiments with solar \cite{solar}, reactor \cite{kl2,chooz},
atmospheric \cite{skatm} and accelerator \cite{k2k,minos}
neutrinos. The existence of neutrino flavor mixing and oscillations
have been established beyond doubt. The \sig{} range
for the mixing parameters governing
solar neutrino oscillations is \footnote{We adopt
the convention in which $m^2_{ij}=m_i^2 - m_j^2$.}
$\ms=(7.2-9.2)
\times 10^{-5}$ eV$^2$ and $\sss=0.25-0.39$ \cite{solglobal},
while the atmospheric neutrino oscillation parameters are
restricted to be within the range $\ma=2.0-3.2 \times 10^{-3}$ and
$0.34<sin^2\theta_{23}<0.68$ with $\sin^22\theta_{23} < 0.9$
\cite{solglobal}.
However, the last mixing angle $\theta_{13}$ remains
unmeasured, although we do know that it must be small. The current \sig{}
upper bound on the allowed
values for this parameter is $\sch < 0.044$ \cite{solglobal}.
If $\theta_{13}$ is close to the upper bound it opens
up the possibility of observing CP violation in the lepton
sector -- the CP phase $\delta_{CP}$ being
the second missing link in our measurement of
the neutrino mass matrix. Also still unknown is
$sgn(\ma)$, which determines the ordering of the neutrino
mass spectrum, {\it aka} the neutrino mass hierarchy. The case
$\ma > 0$ corresponds to the normal mass hierarchy (N3)
while $\ma < 0$ corresponds to the inverted mass
hierarchy (I3).\footnote{Note that the same notation is used
even when the neutrino mass spectrum is quasi-degenerate since
the NH and IH terms here refer only to the $sgn(\ma)$.}
A number of suggestions have been put forward to
determine these three unmeasured parameters. These include
the measurement of antineutrinos from reactors with detector
set-up aiming to achieve sub-percent level in systematic
uncertainties \cite{white}, using
intense conventional $\numu$ beams produced
by decay of accelerated pions \cite{t2k,nova},
using $\nue$ beams produced by decay of accelerated radioactive ions
stored in rings (``Beta-Beams'')
\cite{betabeam} and using $\nue$ and $\numu$
beams produced by decay of accelerated muons stored in rings
(``Neutrino Factory'') \cite{nufact}. Most of these proposed
experiments are expected to be very expensive as well as
technologically challenging.

In principle, information on two of the three parameters mentioned
above can be obtained by observing the neutrinos released
from a galactic supernova. Detailed studies on the potential
of using supernova neutrinos to unravel the mass hierarchy and
$\theta_{13}$ have been performed in the context of
water Cerenkov detectors in \cite{water} and (large) scintillator
detectors in \cite{scin}. It was shown that the neutrino
telescope, IceCube, even though designed to observe ultra high
energy neutrinos, could be used very effectively to detect
low energy supernova neutrinos \cite{raffelticecube,us}.
Most studies on determining of $sgn(\ma)$ and $\theta_{13}$
with supernova neutrinos depend on the hierarchy between the
average energies of $\nue$, $\anue$ and $\nu_x$, where $\nu_x$
stands for $\numu,\anumu,\nutau$, or $\anutau$. However the most
recent supernova models which take into account the full range of the significant neutrino
transport processes, predict very little difference between
the average energies of $\anue$ and $\nu_x$ \cite{snnew}.
At the same time
they also seem to be inconsistent with equi-partition of
energy between the different neutrino species, an assumption
of all earlier papers on supernova neutrinos \cite{totani97}.
In a nutshell,
model uncertainties in the supernova parameters could
washout the oscillation effects and render
this method of hierarchy and $\theta_{13}$ determination
useless.

Less model dependent signatures of $sgn(\ma)$ and
$\theta_{13}$ can be seen through the ``shock effects'' in the
supernova neutrino signal in large detectors
\cite{us, fullerfs,oldfsonly,bargerfs,raffeltforplusrev,
lisiforplusrev}. The steep density profile
at the shock front results in a drastic change in the oscillation
probability and hence can yield information on the
neutrino properties ($sgn(\ma)$ and $\theta_{13}$)
despite the uncertainties associated
with the supernova dynamics and neutrino transport.
However it is also expected that the shock, as it moves outward,
will leave behind a turbulence in the density profile of the
supernova matter. Early studies indicated that this could severely obscure the oscillation signals due to the shock wave \cite{turbbari}. Here
we reexamine this in detail following a recent analysis of the
the effect of turbulence in
\cite{turbfried}. In this the nature of the shock plays an important role. In addition to the initial forward shock  it is expected from supernova
simulations that a reverse shock is formed \cite{raffeltforplusrev}. We find that the effects of the reverse shock are wiped out by turbulence but that the effects of the forward shock, while changed, are still significant and leave a clear signal of the resonant oscillation.

Another important question in neutrino physics
is how many (if any) sterile neutrino species there are.
The first, and so far the only, experimental evidence which
requires the presence of sterile neutrinos comes from
the LSND experiment \cite{lsnd}. While the so-called 2+2 and 3+1
scenarios involving only 1 sterile neutrino \cite{sterileold}
are now comprehensively disfavored by the global neutrino data,
the 3+2 scheme with 2 sterile neutrinos mildly mixed with the
3 active neutrinos has a more acceptable fit to
all data including LSND \cite{threeplustwo} although the value of the associated LSND mixing angle is still problematic  \cite{strumia}.
The LSND result will be
checked by the on-going MiniBOONE experiment \cite{miniboone}
which is expected to
give the final verdict at least on the oscillation
interpretation of the LSND data.  However,
even though MiniBOONE should refute the LSND result,
it will still leave ample room for the existence of extra sterile
neutrinos with mixing angles below its sensitivity limit. We show that measurement of supernovae oscillation effects
provide a much more sensitive probe for sterile neutrinos and analyze in detail the signal that results assuming the 3+2 LSND scenario.

In \cite{us} we analyzed the signature in the IceCube detector of neutrino oscillation occuring within a galactic
supernova,  taking shock effects into account
and assuming that the 3+2 mixing scheme was true. We
considered all possible neutrino mass spectrum with 3 active and
2 sterile neutrinos and showed that IceCube could distinguish
between these different mass spectra. We also showed how well
IceCube could distinguish the case where there were only 3 active
neutrinos from the ones where there are 2 additional
sterile neutrinos. In this paper we repeat this analysis for the case in megaton water \chr detectors, including the effect of turbulence.
Water \chr detectors have a good energy resolution and allow for
the reconstruction of the energy of the incoming neutrinos.
Thus, for such detectors, we will have information on both the time as well as the
energy of the arriving supernova neutrinos.
We calculate the number of supernova $\anue$ events
in a generic megaton water \chr detector. We calculate this
expected signal first for the case of three active neutrinos
and then for the case with two extra sterile species.
We show that
the presence of sterile neutrinos changes the time
evolution of the average energy as well as the flux of the supernova
neutrinos. Therefore, this can be used to indicate the
presence of sterile neutrino and also to
distinguish between the
different possible mass spectrum scenarios. The effect of the shock
is shown to be significant on the resultant signal.
We discuss the issue of turbulence and
take into account the turbulence due to the passage of
shock.

The paper is organized as follows. We begin
in Section 2 with a brief
review of neutrino oscillations inside supernova. We
discuss the effect of the shock wave on the oscillation probability,
with and without the turbulence caused by the passage of shock.
The impact on the neutrino oscillation probability
by the turbulence in the matter density behind the
shock is clearly outlined. In Section 3 we discuss the
neutrino spectrum produced inside the supernova and
their mode of detection in water \chr experiments.
In section 4 we present our results and discuss the implications
in Section 5.
We end with our conclusions in Section 6.

\section{Neutrino Oscillations Inside the Supernova}

When neutrinos propagate in matter, they pick up an extra potential energy
induced by their charged current and neutral current interactions
with the ambient matter \cite{msw1,msw2,msw3}.
Since normal matter only contains
electrons, only $\nue$ and $\anue$ undergo both charged as
well as neutral current interactions, while the other 4
active neutrino species
have only neutral current interactions.
The evolution equation of (anti)neutrinos inside the supernova
can be written in the flavor basis as
\begin{equation}
{\cal H}=\frac{1}{2E}(U{\cal M}^{2}U^{\dag } + {\cal A})~,
\label{eq:schrod}
\end{equation}
where $U$ is a unitary matrix and is defined by $|\nu _{i}\rangle
=\sum_\alpha U_{i\alpha}|\nu _{\alpha }\rangle$, $\nu _{i}$
and $\nu _{\alpha }$ being the mass and flavor eigenstates
respectively.
For the case where we consider two extra sterile neutrinos,
$i=1-5$ and $\alpha $=e, $\mu $, $\tau $, $s_{1}$ or $s_{2}$, where $%
s_{1}$ and $s_{2}$ are the two sterile neutrinos.
The matrices ${\cal M}^{2}$ and ${\cal A}$ are
given respectively by
\begin{equation}
{\cal M}^{2}=Diag(m_{1}^{2},m_{2}^{2},m_{3}^{2},m_{4}^{2},m_{5}^{2})\,,
\end{equation}
\begin{equation}
{\cal A}=Diag(A_{1},0,0,A_{2},A_{2})\,, \label{eq:5mat}
\end{equation}
\begin{equation}
A_1=A_{CC}=\pm \sqrt{2}G_{F} \rho N_A Y_{e}\times2E\,, \label{eq:5ccmat}
\end{equation}
\begin{equation}
A_2 = -A_{NC}=\pm \sqrt{2}G_{F}\rho N_A (1-Y_{e})\times E \,.\label{eq:A2}
\end{equation}%
The quantities $A_{CC}$
and $A_{NC}$ are the matter induced charged current and neutral
current potentials respectively and depend on the Fermi constant
$G_F$, matter density $\rho$, Avagadro number $N_A$,
electron fraction $Y_e$ and energy of the neutrino $E$. The
``$+$'' (``$-$'') sign in Eqs. (\ref{eq:5ccmat})
and (\ref{eq:A2}) corresponds to neutrinos (antineutrinos).

The corresponding expressions for the three neutrino framework
follows simply by dropping the extra 2 states due to the sterile
components. Note that we have recast the matter induced
mass matrix such that the
neutral current part $A_{NC}$, which appears for all the
three active flavors, is filtered out from the first three diagonal
terms and hence it appears as a negative
matter potential for the sterile states which do not
have any weak interactions. Therefore, for the three active
neutrino set-up the mass matrix is
\begin{equation}
{\cal M}^{2}=Diag(m_{1}^{2},m_{2}^{2},m_{3}^{2})\,,
\end{equation}
\begin{equation}
{\cal A}=Diag(A_{1},0,0)\,. \label{eq:3mat}
\end{equation}
The mixing matrix $U$ is given in terms of mixing angles and
CP-violating phases. If CP conservation is assumed the mixing matrix
takes the form
\begin{equation}
U=\prod_{B>A}^{n}\prod_{A=1}^{n-1}R^{AB(\theta_{AB})}~,
\end{equation}%
where $n=3$ for only three active neutrinos and $n=5$
when two additional sterile neutrinos are present and
$R^{AB}$ is an $n\times n$ rotation matrix about the AB plane.

Assuming that all phases get averaged out, the survival probability
$\bar P_{ee}$ of $\anue$ after they have propagated through the
supernova matter can be written as\footnote{Since we are interested
in the detection of supernova neutrino in water \chr detectors where
the largest number of events are expected through the capture of
$\anue$ by protons, we will mostly refer to the survival probability
of $\anue$ ($\bar P_{ee}$) in this section. However, similar
expressions are valid even for the $\nue$ survival probability.} \be
\bar P_{ee} = \sum_{i,j} |U_{ej}|^2|U_{ei}^m|^2P_{ij}, \ee where,
$P_{ij}=|\langle\bar\nu_j|\bar\nu_i^m\rangle |^2$ and ~$U_{ei}$ and
$U_{ei}^m$ are the elements of the mixing matrix in vacuum and
inside matter respectively and
$|\langle\bar\nu_j|\bar\nu_i^m\rangle|^2$ is the effective ``level
crossing'' probability that an antineutrino state created as
$|\bar\nu_i^m\rangle$ inside the supernova core emerges as the state
$|\bar\nu_i\rangle$ in vacuum. Largest flavor conversions occur at
the resonance densities. For two flavor oscillations, the resonance
condition for antineutrinos involving only either the $\anue$ or the
sterile states is given by \cite{msw2}
\begin{equation}
|A_{k}|=(-1)^k\Delta m^{2}_{ji}\cos 2\theta_{ij}  ~, \label{eq:res}
\end{equation}%
where $k$ could be either 1 or 2,
$\Delta m^2_{ji}\equiv m_j^2 - m_i^2$ is the mass squared
difference and $\theta_{ij}$ is the mixing angle between the two states
in vacuum. Note in Eq. (\ref{eq:res}), we included the sign factor
$(-1)^k$ in order to take into account the fact that $A_{CC}<0$ and
$A_{NC} >0$ for the antineutrinos.
Since $A_{CC}<0$ ($A_{NC} > 0$), the
condition of resonance is satisfied only when the relevant
$\Delta m^2_{ji} < 0$ ($\Delta m^2_{ji} > 0$). For the resonance
between the $\anue$ and sterile states one must satisfy the
condition
\be
A_{CC}-A_{NC}=\Delta m^{2}_{ji}\cos 2\theta_{ij}  ~, \label{eq:res2}
\ee
which is satisfied for antineutrinos when $\Delta m^2_{ji} < 0$
since $A_{CC} > A_{NC}$.

The probability of
level crossing from one mass eigenstate to another
is predominantly non-zero only at the resonance.
In the approximation
that the individual two flavor resonances are far apart, the
effective level crossing probability can be written in
terms of the individual two flavor ``flip probability'' $P_{ij}$.
The ``flip probability''
between the
two mass eigenstates
used in this paper is given by \cite{petcovflip}
%
\begin{equation}
P_{ij}=\frac{\exp (-\gamma \sin ^{2}\theta_{ij} )-\exp (-\gamma )}{1-\exp
(-\gamma )}~,
\label{eq:jumpprob}
\end{equation}
\begin{equation}
\gamma =\pi \frac{\Delta m^{2}_{ji}}{E}\left\vert \frac{d\ln
A}{dr}\right\vert _{r=r_{mva}}^{-1}~,
\end{equation}%
where $A$ is $A_{CC}$, $A_{NC}$ or $A_{CC}-A_{NC}$ depending
on the states involved in the resonance,
$r_{mva}$ is the position of the maximum violation of
adiabaticity
($mva$) \cite{mva} and is defined as%
\begin{equation}
A(r_{mva})=\Delta m^{2}_{ji}~.
\end{equation}
One can see from the above equations that $P_{ij}$ and
hence the transition probability depends crucially
both on the mixing angle as well as on the density
gradient.
If for given $\Delta m^2_{ji}$ and $\theta_{ij}$, the
density gradient is small enough so that $\gamma \gg 1$ and $\gamma
\sin^2\theta \gg 1$, $P_{ij} \simeq 0$ and the transition is
called adiabatic. On the other hand, if the density gradient is
very large or $\theta_{ij}$ is very small, then
$P_{ij}\simeq \cos ^{2}\theta_{ij}$.
In this case we have an extreme non-adiabatic
transition. For all intermediate values of the
density gradient and $\theta_{ij}$,
$P_{ij}$ ranges between [0-1].

\subsection{Neutrino Oscillations with Static Density Profile}

\begin{table}[p]
\begin{center}
\begin{tabular}{|c|c|c|c|}
\hline
&&& \\
Hierarchy & i & $\bar P^{m}_{ei}$ & $\bar P^{m}_{xi}$ \\
&&& \\\hline\hline
N2+N3 & 1 & 1 & 0 \\

& 2 & 0 & $P_{25}P_{24}$ \\
& 3 & 0 & $%
P_{25}(1-P_{24})(1-P_{34})+(1-P_{25})(1-P_{35})P_{34}+P_{35}P_{34} $ \\
& 4 & 0 & $%
P_{25}(1-P_{24})P_{34}+(1-P_{25})(1-P_{35})(1-P_{34})+P_{35}(1-P_{34})$ \\
& 5 & 0 & $(1-P_{35})+(1-P_{25})P_{35}$ \\ \hline
N2+I3 & 1 & $P_{13}$ & $P_{35}P_{34}(1-P_{13})$ \\
& 2 & 0 & $%
P_{35}(1-P_{34})(1-P_{24})+(1-P_{35})(1-P_{25})P_{24}+P_{25}P_{24} $ \\
& 3 & $1-P_{13}$ & $P_{35}P_{34}P_{13}$ \\
& 4 & 0 & $%
P_{35}(1-P_{34})P_{24}+(1-P_{35})(1-P_{25})(1-P_{24})+P_{25}(1-P_{24})$ \\
& 5 & 0 & $(1-P_{25})+(1-P_{35})P_{25}$ \\ \hline
H2+N3 & 1 & $P_{14}$ & 0 \\
& 2 & 0 & $P_{25}$ \\
& 3 & 0 & $P_{35}+(1-P_{25})(1-P_{35})$ \\
& 4 & $1-P_{14}$ & 0 \\
& 5 & 0 & $(1-P_{35})+(1-P_{25})P_{35}$ \\ \hline
H2+I3 & 1 & $P_{14}P_{13}$ & $P_{35}(1-P_{13})$ \\
& 2 & 0 & $(1-P_{35})(1-P_{25})+P_{25}$ \\
& 3 & $(1-P_{13})P_{14}$ & $P_{35}P_{13}$ \\
& 4 & $(1-P_{14})$ & 0 \\
& 5 & 0 & $(1-P_{25})+(1-P_{35})P_{25}$ \\ \hline
I2+N3 & 1 & $P_{15}P_{14}$ & 0 \\
& 2 & 0 & 1 \\
& 3 & 0 & 1 \\
& 4 & $P_{15}(1-P_{14})$ & 0 \\
& 5 & $1-P_{15}$ & 0 \\ \hline
I2+I3 & 1 & $P_{15}P_{14}P_{13}$ & $1-P_{13}$ \\
& 2 & 0 & 1 \\
& 3 & $P_{15}P_{14}(1-P_{13})$ & $P_{13}$ \\
& 4 & $P_{15}(1-P_{14})$ & 0 \\
& 5 & $1-P_{15}$ & 0 \\
&  &  &  \\ \hline
\end{tabular}%
\end{center}
\caption{\label{tab:jumpprob}The probabilities $\bar P_{ei}^m$ and
$\bar P_{xi}^m$ for three active neutrinos plus two sterile neutrinos,
where $\bar P_{\protect\alpha i}^{m}$ is given by Eq. (\ref{eq:flipprob})
and $P_{ij}$ is the flip probability at the resonance between the
$\protect\nu _{i}$ and $\protect\nu _{j}$ mass eigenstates.}
\label{flipprobthreeplustwo}
\end{table}

{\bf{Three active neutrinos only:}}
As discussed above,
in the three neutrino framework we have two possibilities
for the neutrino mass spectrum,
the normal ($\ma >0$) and inverted ($\ma<0$) hierarchy.
We will call them N3 and I3 cases respectively.
Also, since here
we have 2 independent $\Delta m^2_{ji}$, we can have
2 resonances and hence the final level crossing probability
will be given in terms of the 2 individual flip probabilities.
However, it is now known at more than $6\sigma$ C.L.
from the solar neutrino data that $\ms >0$, and hence the
$\ms$ driven resonance condition given by Eq. (\ref{eq:res})
is satisfied only for the neutrino channel.
But the sign of $\ma$ is still unknown. Hence the $\ma$
driven resonance can appear in either the neutrino or
antineutrino sector depending on whether the mass hierarchy
turns out to be normal or inverted respectively.
Therefore, for $\ma <0$, the $\anue$
survival probability (neglecting earth
matter effects) is given by
\begin{equation}
\bar P_{ee}=P_{13}|U_{e1}|^{2}+(1-P_{13})|U_{e3}|^{2}
\label{eq:3nupee}
\end{equation}%
where $P_{13}$ is the flip probability between the 1 and 3 states
and can be calculated using Eq. (\ref{eq:jumpprob}) with
$\Delta m^2_{ji} \equiv \ma$ and $\theta_{ij} \equiv \theta_{13}$.
For $\ma >0$ the resonance does not occur in the
antineutrino channel and therefore
\be
\bar P_{ee}= |U_{e1}|^2
\ee

\noindent
{\bf{Three active plus two sterile neutrinos:}}
For the three active and
two sterile neutrino framework we
have 4 independent mass squared differences and
hence can have as many as six
possibilities for the mass spectrum which we call \cite{us}:
\be
{\rm N2+N3:}~ &\ma>0&, \Delta m^2_{41}>0 ~{\rm and}~ \Delta m^2_{51}>0~, \label{eq:n2n3}\\
{\rm N2+I3:}~ &\ma<0&, \Delta m^2_{41}>0 ~{\rm and}~ \Delta m^2_{51}>0~, \label{eq:n2i3}\\
{\rm H2+N3:}~ &\ma>0&, \Delta m^2_{41}>0 ~{\rm and}~ \Delta m^2_{51}<0~, \label{eq:h2n3}\\
{\rm H2+I3:}~ &\ma<0&, \Delta m^2_{41}>0 ~{\rm and}~ \Delta m^2_{51}<0~, \label{eq:h2i3}\\
{\rm I2+N3:}~ &\ma>0&, \Delta m^2_{41}<0 ~{\rm and}~ \Delta m^2_{51}<0~, \label{eq:t2n3}\\
{\rm I2+I3:}~ &\ma<0&, \Delta m^2_{41}<0 ~{\rm and}~ \Delta
m^2_{51}<0~,\label{eq:i2i3} \ee with $\ms>0$ always. Since we have 4
different independent $\Delta m^2_{ji}$ and both active and sterile
components, we can have many more resonances in this case. In fact,
we expect to have resonances between the 1-2, 1-3, 1-4 and 1-5
states due to the charged current term and between 2-4, 2-5, 3-4 and
3-5 states due to the neutral current term. Again since $\ms>0$, the
1-2 resonance happens only for neutrinos, while the other resonances
can happen in either the neutrino or antineutrino channel depending
on whether the corresponding mass squared difference is positive or
negative. Another difference between the only active and active plus
sterile cases is that while for only active neutrino cases it was
enough to compute just the survival probability $\apee$, with
sterile neutrinos in the fray, the total neutrino flux produced
inside the supernova does not remain constant as the active flavors
also  disappear into sterile species. Thus we will need to calculate
the probabilities $\apee$ as well as $\apxe$. The oscillation
probability in general is given by
\begin{equation}
\bar P_{\alpha \beta }=\sum_{i}\bar P_{\alpha i}^{m} \bar P_{i\beta
}^{\oplus }~,\label{eq:genflipprob}
\end{equation}
where $P_{\alpha i}^{m}$ and $P_{i\beta }^{\oplus }$ are
respectively the
$\bar\nu_\alpha \rightarrow \bar\nu_i^m$ transition probability
in the supernova and
$\bar\nu_i^\oplus \rightarrow \bar\nu_\beta$  transition probability
in the earth.
In the absence of earth matter effects
\be
\bar P_{ie }^{\oplus } = |U_{ei}|^2
\ee
and
\begin{equation}
\bar P_{\alpha i}^{m}=\sum_{j}|U_{\alpha j}^{m}|^{2}P_{i j}~,
\label{eq:flipprob}
\end{equation}
where $P_{ij}$
is the flip probability between the $i$ and $j$ mass eigenstates,
given by Eq. (\ref{eq:jumpprob}).
We explicitly give the probabilities $\bar P_{ei}^m$ and
$\bar P_{xi}^m$ for all the possible mass spectra in
Table \ref{flipprobthreeplustwo}.
For further details and the level crossing diagrams for each of these
mass spectra, we refer the reader to \cite{us}.

\subsection{Effect of the Shock Wave on $\bar P_{ee}$}

\begin{figure}[h]
\begin{center}
\includegraphics[width=17cm]{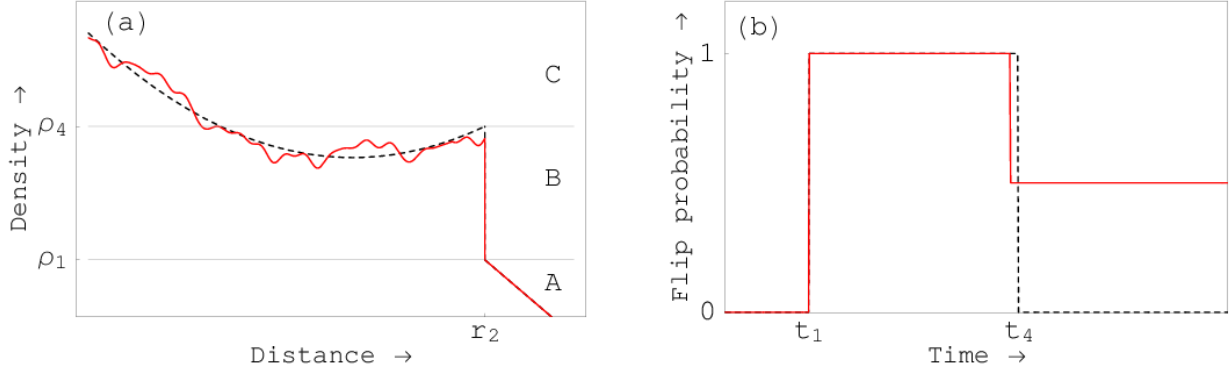}
\end{center}
\caption{(a) The density profile of the forward shock as a function
of the distance from the core of the supernova. (b) The resulting
flip probability as a function of time. The dashed black
line is for the case of no turbulence and the solid red line includes
the effect of turbulence.}
\label{fig:forward}
\end{figure}

\begin{figure}[h]
\begin{center}
\includegraphics[width=17cm]{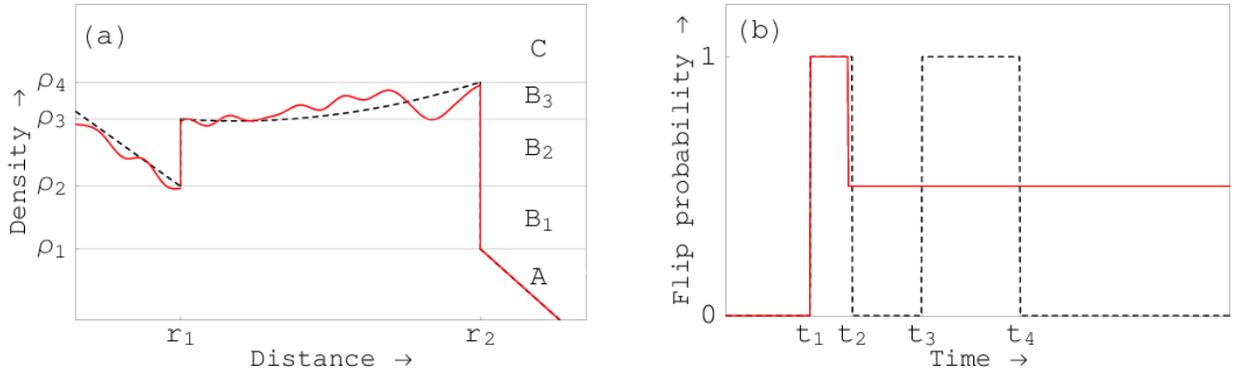}
\end{center}
\caption{(a) The density profile of the forward and reverse shock as
a function of the distance from the core of the supernova. (b) The
resulting flip probability as a function of time. The dashed black
line is for the case of no turbulence and the solid red line includes
the effect of turbulence.} \label{fig:forwardandback}
\end{figure}

It is believed that when the core of a collapsing supernova reaches
nuclear density the collapse rebounds forming a strong outward
shock. The shock stalls and is eventually regenerated via neutrino
heating. Numerical simulations show that as well as a forward shock
a reverse shock forms and the region behind the shock wave is highly
turbulent \cite{raffeltforplusrev}.
The detailed density profiles from numerical simulations are
not available to us, therefore we consider the
simplified profile for the shock wave used in
\cite{oldfsonly,lisiforplusrev}. A schematic diagram showing
the density change at the shock front for forward shock and
for forward and reverse shock is shown
by the dashed black lines in
Figs. \ref{fig:forward}(a) and \ref{fig:forwardandback}(a).

As can be seen from these figures,
the effect of the
supernova shock wave is to generate very sharp
changes in the density gradient, which
is expected to change the flip probability.
Since, as mentioned above, the flip probability is
calculated at the position of resonance, the effect of the
shock will be to modify the flip probability and hence
$\bar P_{ee}$
when the shock crosses the
resonance density for a certain $\Delta m^2_{ji}$.
Also, we see from the figures that the shock
generates fluctuations in the density
gradient. This results in the same $\Delta m^2_{ji}$
producing multiple resonances which are relatively close together.
If we assume that the phase effects can be neglected even in this
case \cite{phase}, then the individual
resonances can be considered as independent two generation
resonances and the net flip probability $P_H$ can be expressed in
terms of the multiple flip probabilities $P_i$ as
\cite{kuopanta,oldfsonly}
\be
\begin{pmatrix}
1-P_H & P_H\\
P_H & 1-P_H
\end{pmatrix}
= \prod_{i=1,n}
\begin{pmatrix}
1-P_i & P_i\\
P_i & 1-P_i
\end{pmatrix}~,
\label{eq:multiplePJ}
\ee
where $n$ is the number of resonances
occurring for the same $\Delta m^2_{ji}$ due to the shock effect.
In particular, if the mixing angles are sufficiently large
then in the region of the supernova where the density can be
approximated as almost static,
the flip probability between the mass eigenstates is approximately
zero (the resonance is adiabatic). As the shock wave passes through
the resonant density, the flip probability increases to
approximately one (the resonance becomes
strongly non-adiabatic). As the shock passes over, then
in the approximation that there is no turbulence (which will
be considered in the next subsection), the flip probability
goes back to zero.
The resultant
flip probability has a time dependence shown in Figs.
\ref{fig:forward}(b) and \ref{fig:forwardandback}(b).

Since
the resonance densities are different for the different
$\Delta m^2_{ji}$ involved, the shock wave will cross them at
different times.
Therefore, one can follow the evolution of
the shock wave by following the time profile of the
supernova signal \cite{us}. Also, since the
resonance density is determined by the energy of the
neutrino, different energy neutrinos have their resonance
position as different points and thus get affected by the
shock at different times. Thus the shock imprints its
signature on both the energy spectrum as well as the time
profile of the neutrinos arriving on earth.
The effect of the shock wave is discussed in
\cite{fullerfs,oldfsonly} for the forward shock and in
\cite{raffeltforplusrev,lisiforplusrev}
when there is a reverse shock as well.
All these papers used large/megaton water detectors
for the supernova neutrino signal on earth.
The possibility of using shock waves to discern the existence of
eV mass sterile neutrinos in the supernova signal in IceCube
was discussed in \cite{us}.

\subsection{Turbulence}
Numerical simulations have shown that a highly turbulent region
forms behind the shock wave \cite{numturb}. The fluctuations in
these turbulence cause flips between the mass eigenstates
\cite{turbbari,turbfried}. The length scale of the fluctuations for which the
flips are dominant are at the length scale

\begin{equation}
L_{fluct}=\frac{4 \pi E}{\Delta m^{2} \sin2\theta}
\end{equation}

For the atmospheric resonance
$L_{fluct}=47km\frac{E}{15MeV}\frac{0.3}{\sin2 \theta_{13}}$, which
is smaller than the resolution of the numerical simulations and
therefore a physical model of the fluctuations is required.
Analytical solutions for a density fluctuating medium have only been
found for the un-physical delta-correlated noise

\begin{equation}
\langle\delta n(x)\delta n(y)\rangle=2L_{0}\delta n ^{2}\delta(x-y)
\end{equation}
where the fluctuations are fully correlated for length scales less
than $L_{0}$ and $\delta n ^{2}$ is the average squared fluctuation.
Simulations using these analytical solutions show that the system
becomes depolarized for large fluctuations, the flip probability
$P_{turb}\sim0.5$.

A more physically reasonable model of turblence is provided by the Kolmogorov spectrum in which the
large scale density fluctuations cascade into density fluctuations
on a small scale. These density fluctuations exhibit a power-law
spectrum

\begin{equation}
\int dx\langle\delta n(0)\delta n(x)\rangle e^{-\textit{i}k
x}=C_{0}k^{\alpha}
\end{equation}
where $n(x)$ is the density of the supernova at the position $x$,
$k=2\pi/\lambda$ where $\lambda$ is the wavelength of the
fluctuation, $C_0$ is a constant and $\alpha=-5/3$. The density
fluctuations on a large scale (the size of the shock wave) can be taken from
numerical simulations and used to determine the value of $C_{0}$.
Then this can be used to estimate the size of the fluctuations at
the scale at which dominate the oscillations.
Numerical solutions for the case of Kolmogorov
turbulence also show that the system becomes depolarized for large
turbulence \cite{turbfried}. Currently no exact solutions exist for
Kolmogorov turbulence but the flip probability can be calculated
in the perturbative limit giving

\begin{equation}
P_{turb}\simeq\left\{
\begin{array}{c c c}
  \frac{0.84G_{F}C_{0}}{\sqrt{2}|n_{0}'|}(\frac{\Delta m^{2}\sin2\theta}{2E})^{-2/3}
   & for & \frac{\pi\left(\Delta m^{2}\sin2\theta\right)^{2}}{4E|A'|}\gg1 \\
  1 & for & \frac{\pi\left(\Delta m^{2}\sin2\theta\right)^{2}}{4E|A'|}\ll1
\end{array}
 \right.
\end{equation}

This is in the perturbative limit but a depolarization criterion can
be defined by $P_{turb}\sim0.5$. Fluctuations
which satisfy this condition are expected to depolarize the neutrino
flavours which are maximally mixed at the resonance. The current
simulations show that this criteria is satisfied by a large margin for both the
atmospheric and sterile resonances. In these cases the resonant neutrinos are expected to be depolarized in the turbulent region behind the shock wave.

The flip
probability of the density profile including the effect of
turbulence is shown in Figs. \ref{fig:forward} and
\ref{fig:forwardandback}. Note that for the forward shock the initial rise is unchanged by turbulence but the fall is reduced by the turbulent depolarisation. This just reflects the fact that the turbulence follows the shock wave. The effect on the reverse shock is dramatic, the depolarisation wiping out the reverse shock structure. Note too that the depolarisation following the shock front can lead to interesting observable effects in active neutrino oscillation for the case there are additional sterile neutrinos. This is because the active-sterile oscillation will destroy the near degeneracy between $\bar{\nu_e}$ and $\nu_x$. As a result the subsequent atmospheric neutrino oscillation signal, which vanishes if the active neutrino luminosities and spectra are equal, can be significantly enhanced. We will demonstrate these effects in the subsequent sections.

\section{Observing Supernova Neutrinos in Water}

\subsection{Energy Spectra of Supernova Neutrinos}

In a type-II supernova approximately $3\times10^{53}$ ergs of energy
is released and about 99\% of this is in the form of neutrinos. The
energy spectra of these neutrinos is determined by their
interactions with matter. As the effect of weak magnetism in
muon or tau production can be neglected, the spectra of
$\nu_{\mu},\bar{\nu}_{\mu}, \nu_{\tau}$ and $\bar{\nu}_{\tau}$ are
approximately equal and are collectively denoted by $\nu_{x}$.
The
initial spectra of neutrino species $\nu_\alpha$ from a supernova is
parameterized as \cite{spectra}
\begin{equation}
F^{0}(E,t)=
\frac{\Phi(t)}{\langle E \rangle(t) } \frac{(\alpha(t)+1
)^{\alpha(t)+1}}{\Gamma (\alpha(t) +1)} \left( \frac{E}{\langle
E\rangle(t) }\right) ^{\alpha(t) }\exp \left( -(\alpha(t)
+1)\frac{E}{\langle E \rangle(t) }\right) \label{eq:flux}
\end{equation}
where
$\langle E\rangle $ and $\Phi$ are the average energy and
total number flux
and $\alpha$ is a dimensionless parameter which typically takes the
values 2.5-5. For the results presented in this paper, we have
assumed $\alpha_{\bar e} = 3 $ and $\alpha_x = 4 $.

To date only the Lawrence Livermore (LL)
group \cite{totani97} has published
detailed results of the energy spectra of neutrinos for the duration
of the supernova.
In the LL simulations the luminosities of all flavours of
neutrinos are approximately equal for times post bounce $\gtrsim
0.1$ s. However there is a distinct difference in the average energies.
This is due to the additional charged current interactions of the
$\nu_{e}$ and $\bar{\nu}_{e}$ with ambient matter.
However these simulations did not
include all the neutrino interactions now believed to be
important \cite{snnew}.
Ongoing work by the Garching group
including all possibly significant interactions predict the neutrino fluxes and average
energies labeled G1 and G2 in Table \ref{tab:enlmodel}.
This shows that the average
energies between neutrino flavours become very similar and the
luminosities are no longer equal.
In order to
compare the impact of the uncertainties on the average energies and
fluxes of the neutrinos obtained in different supernova computer
simulations, we will present our results using supernova neutrino
parameters given by both the Lawrence Livermore and Garching groups.
Specifically, we consider the three cases shown in Table
\ref{tab:enlmodel} \cite{raffeltforplusrev}.
For further
details,
we refer the reader to our earlier paper \cite{us}.


\begin{table}
\begin{center}
\begin{tabular}{|c|c|c|c|c|c|}
\hline Model & $\langle E^0_{\nue} \rangle$ & $\langle E^0_{\anue}
\rangle$  & $\langle E^0_{\nu_x} \rangle$ &
$\frac{\Phi^0_{\nue}}{\Phi^0_{\nu_x}}$ &
$\frac{\Phi^0_{\anue}}{\Phi^0_{\nu_x}}$\cr \hline LL & 12 & 15 & 24
& 2.0 & 1.6 \cr G1 & 12 & 15 & 18 & 0.8 & 0.8 \cr G2 & 12 & 15 & 15
& 0.5 & 0.5 \cr \hline
\end{tabular}%
\end{center}
\caption{\label{tab:enlmodel} The average energies and total fluxes
characterizing the primary neutrino spectra produced inside the
supernova. The numbers obtained in the Lawrence Livermore
simulations are denoted as LL, while those obtained by the Garching
group are denoted as G1 and G2. }
\end{table}

\subsection{Signal in Water \chr Detectors}

Supernova neutrinos with energy in the MeV regime will be dominantly
detected in water through the capture of $\anue$ on
protons\footnote{Water \chr detectors have other detection channels
whereby they can observe $\nue$ and $\nu_x$ (electron scattering and
charged and neutral current interactions on $^{16}O$). However, the
cross-section for these processes are much smaller and hence they
are not considered here.} \be \bar{\nu}_{e} + p\rightarrow n +
e^{+}~. \ee The emitted positron will be observed in the detector
and its energy and time measured. Hence we should be able to get a
fairly good reconstruction of the incoming $\anue$ energy spectrum
and time profile. Neglecting the recoil of the neutron the energy of
the neutrino $E$ is related to the energy of the positron through
the relation $E_e = E - 1.29$, where all energies are in MeV. The
number of events expected in a water \chr detector from a galactic
supernova explosion is given by \be N= \frac{N_T}{4\pi D^2} \int_{0}^{\infty}
\int_{0}^{\infty} F(E) * \sigma(E) * \varepsilon(E)
* R(E-1.29,E_e)~ dE~ dE_e \ee where $D$ is the distance of the supernova from earth, $N_T$ is the number of
target nucleons in 1 megaton of water, E is the energy of the
neutrino, $E_e$ is the measured energy of the positron, F(E) is the
flux at the detector, as defined in Eqs. (30) and (37), $\sigma(E)$
is the cross section, $\varepsilon(E)$ is the efficiency of
detection and $R(E-1.29,E_e)$ is the energy resolution function. The
efficiency is assumed to be perfect above 7 MeV and vanishing below
this energy. The energy resolution function for which we assume a
Gaussian form \be R(E_T,E_e) =
\frac{1}{\sqrt{2\pi}\sigma_E}\exp\left(\frac{-(E_T-E_e)^2}
{2\sigma_{E}^2}\right)~, \label{eq:resol} \ee where $E_T$ and $E_e$
are respectively the true and measured energy of the positron and we
take the HWHM $\sigma_E(E_T) = \sqrt{E_0 E_T}$, where $E_0 =
0.22$MeV. This is the same efficiency and energy resolution as used
in \cite{raffeltforplusrev, Tomas:2003xn}. Water \chr detectors
usually are expected to have time resolution which is of the order
of nanosecond. In what follows, we will bin our data either in time
bins of 100 ms (at later times) or 10 ms (at earlier times). Since
we expect the time resolution of the detector to be at least 3-4
orders of magnitude better, we do not include any time resolution
function in our calculation of the number of events.

In addition to the total number of events, we also calculate the the
average energy of the detected positrons through the expression \be
\langle E_e \rangle = \frac{\int_{0}^{\infty}\int_{0}^{\infty} E_A *
F(E) * \sigma(E) * \varepsilon(E) * R(E-1.29,E_A)~ dE~ dE_A}
               {\int_{0}^{\infty}\int_{0}^{\infty} F(E) * \sigma(E) * \varepsilon(E) * R(E-1.29,E_A)~ dE~ dE_A}
\ee

We show both the number of events as well as the average energy as a
function of time. We also show the statistical uncertainties
expected in 1 megaton water \chr detectors. The statistical error in
the total number of events are estimated as \be \sigma_{N} & = &
\sqrt{N}~, \ee while that in the average energy is calculated as \be
\sigma_{\langle E\rangle} & = & \sqrt{\frac{\langle
E_e^{2}\rangle-\langle E_e\rangle^{2}}{N}}~, \ee where
$\sigma_{\langle E\rangle}$ is the error in  the average energy, N
is the number of events, $\langle E\rangle$ is the average energy
and $\langle E^{2}\rangle$ is the average energy squared.

The neutrino flux in the detector is
\begin{equation}
F_{\beta }=\sum_{\alpha }F_{\alpha }^{0}P_{\alpha \beta }
\label{eq:fluxprob}
\end{equation}
where $P_{\alpha \beta }$ is the oscillation probability and is
given in Eqs. (\ref{eq:genflipprob})-(\ref{eq:flipprob}) and
$F_{\alpha }^{0}$ is the initial flux of $\nu_{\alpha}$ given in
Eqn. (\ref{eq:flux}).

\subsection{Input Supernova and Oscillation Parameters}

In what follows, we will present results for the typical
values for the fluxes and average energies given in
Table \ref{tab:enlmodel} for the LL, G1 and G2 ``models''.
For the neutrino oscillation parameters, we assume the
best-fit values $\ms=8\times 10^{-5}$ eV$^2$, $\sss=0.31$
and $\ma=2.5\times 10^{-3}$ \cite{solglobal}.
For $\theta_{13}$, we assume that it is
large enough so that away from the shock,
the $\ma$ driven resonant transition
is fully adiabatic.
Typically, this would be satisfied for $\sch \gtap 10^{-3}$.

The other oscillation parameters relevant for supernova
neutrino oscillations in the 3+2 scenario are constrained
by the combined data from
Bugey, CHOOZ, CCFR84, CDHS, KARMEN, NOMAD, and LSND
\cite{lsnd,sbl}.
If we restrict the mass squared differences to lie in the
sub-eV regime then the best-fit comes at
$\Delta m^2_{41}=0.46$ eV$^2$, $\Delta m^2_{51}=0.89$ eV$^2$,
$U_{e4}=0.09$, $U_{e5}=0.125$, $U_{\mu 4}=0.226$ and $U_{\mu 5}=0.16$
\cite{threeplustwo}. These values of the oscillation
parameters give fully adiabatic transition at the
resonance in the supernova when the shock is not present.

\subsubsection{Sterile neutrino sensitivity}
In the detailed estimates presented below we use the sterile neutrino parameters
consistent with an explanation of the LSND experiment. However it is important to stress
that the sensitivity to sterile neutrinos is much better than is needed to probe the LSND range and
that, even if MiniBoone should rule out the sterile mixing angle regime used in our estimates, supernovae
neutrino signals will still be important in the search for evidence for sterile neutrinos. It is straightforward
to quantify the range of sensitivity.
To observe the time dependent effects in the signal due to sterile
neutrinos, the adiabaticity of the sterile resonances needs to be
changed by the shock wave and/or the turbulence. For mixing angles
$\sin^{2}\theta_{ij}\lesssim 4\times10^{-6}$ the sterile resonances
would be non-adiabatic for the entire time of interest. As a result
there would be no oscillations into sterile neutrinos and the signal
would be equivalent to that of only 3 active neutrinos. For
$\sin^{2}\theta_{ij}\gtrsim4\times10^{-4}$ the resonance would be
adiabatic in the regions behind (with no turbulence) and in front of
the shock wave. In this range the active-sterile resonant mixig effects are
measureable. The situation is summarised in Table
\ref{tab:Miniboone}.

\begin{table}[tbp] \centering%
\begin{tabular}{|c|c|}
  \hline
  $\sin^{2}\theta_{ij}$ & Region\\
  \hline
  $6 \times 10^{-3}\lesssim\sin^{2}\theta_{ij}$ & Sensitivity range of MiniBoone\\
  $4\times10^{-4}\lesssim\sin^{2}\theta_{ij}$ & Maximal effect of shock \\
  $4\times10^{-6}\lesssim\sin^{2}\theta_{ij}\lesssim4\times10^{-4}$ & Transition region \\
  $\sin^{2}\theta_{ij}\lesssim4\times10^{-6}$ & No effect of shock\\
  \hline
\end{tabular}
\caption{\label{tab:Miniboone}
The effect of the shock wave for the parameter space of
$\sin^{2}\theta_{ij}$ with i=1,2 and j=3-5.}
\end{table}

 For $4\times10^{-6}<\sin^{2}\theta_{ij}<4\times10^{-4}$
there is a ``transition region'', in which the effects
discussed in this paper
could be observed but may be less prominent. To quantify this first note that
within this region of parameter space
the mixing angles are sufficiently large such that each resonance is
adiabatic in the absence of the shock wave. As well as changing the
adiabaticity of each resonance the mixing angles change the relative
proportion of each flavour eigenstate in each mass eigenstate, as a
result changing the $\bar{\nu}_{e}$ flux, $F_{e}$ in the detector.
The approximate $F_{e}$ for small sterile mixing angles is shown in
Table 4. If the sterile mixing angles and $\theta_{13}$ are scaled
as $\theta_{ij}\rightarrow k\theta_{ij}$, where i=1,2, and j=3-5,
and $k<1$, $F_{e}$ is approximately unchanged except for the N2+I3
and H2+N3 mass hierarchies. If the flux is unchanged both the number
of events and the average energies are unchanged. For the
cases of N2+I3 and H2+N3 mass hierarchies $F_{no shock}$ is scaled as
$F_{no shock}\rightarrow k^{2} F_{no shock}$. As a result the flux
and therefore number of observed events before and after the
propagation of a shock wave scale as $k^{2}$, the statistical
uncertainties scale as $k^{-1}$ and the average
energies remain unchanged. Therefore for $k<1$ the number of
events decreases, and the uncertainties in the number of events and
average energies increases. As the shock wave passes through the
resonance the flux is $F_{e}\simeq F_{shock}$, which is independent
of $k$. Therefore the total number of events is approximately
unchanged. However, as the shock passes through a resonance the
number of events increases, these corresponds to the lowest energy
neutrinos, as a result the average energy decreases. For smaller $k$
the relative increase in the number of events is larger and
therefore the decrease in the average energy is larger. At later
times there is an increase in the average energy as the resonance
condition is satisfied for higher energy neutrinos, for smaller $k$
the relative increase is larger and therefore the increase in the
average energy is larger. As a result the structure of the average
energy plot remains but is stretched for smaller $k$. Simulations
show that the number of events during the shock propagation is
approximately independent of $k$ for $k\lesssim1$ and the average
energy plot is stretched as described above.

\subsection{Three active neutrinos}

\begin{figure}[h]
\begin{center}
\includegraphics[width=17cm]{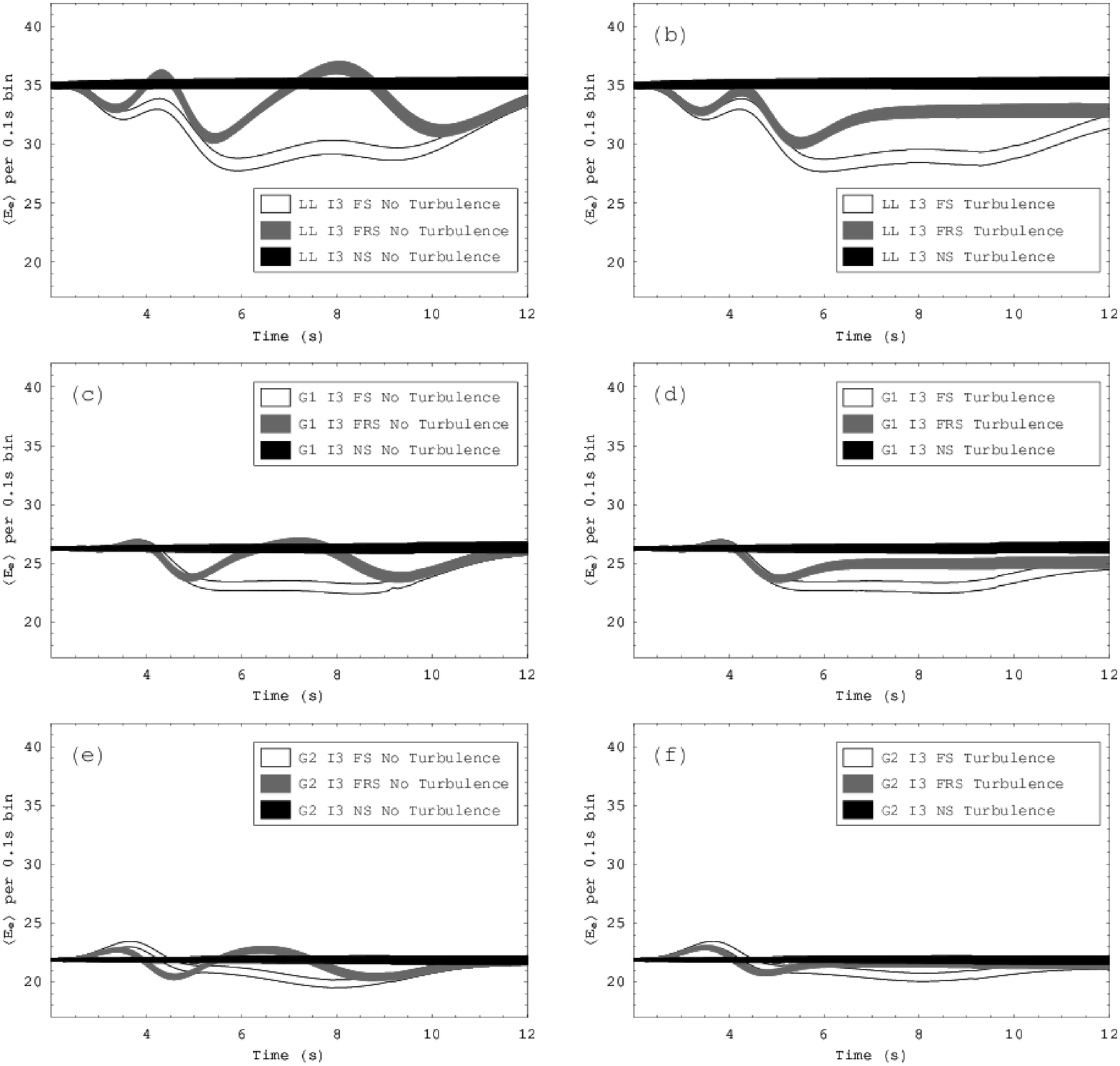}
\end{center}
\caption{The average energy in 100ms bins, for the cases of a
forward shock (FS), a forward and reverse shock (FRS) and no shock
(NS): (a)LL no turbulence, (b)LL with Turbulence, (c)G1 no
turbulence, (d) G1 with turbulence, (e)G2 no
turbulence, (f) G2 with turbulence.}\label{fig:3avenergy}
\end{figure}

\begin{figure}[h]
\begin{center}
\includegraphics[width=17cm]{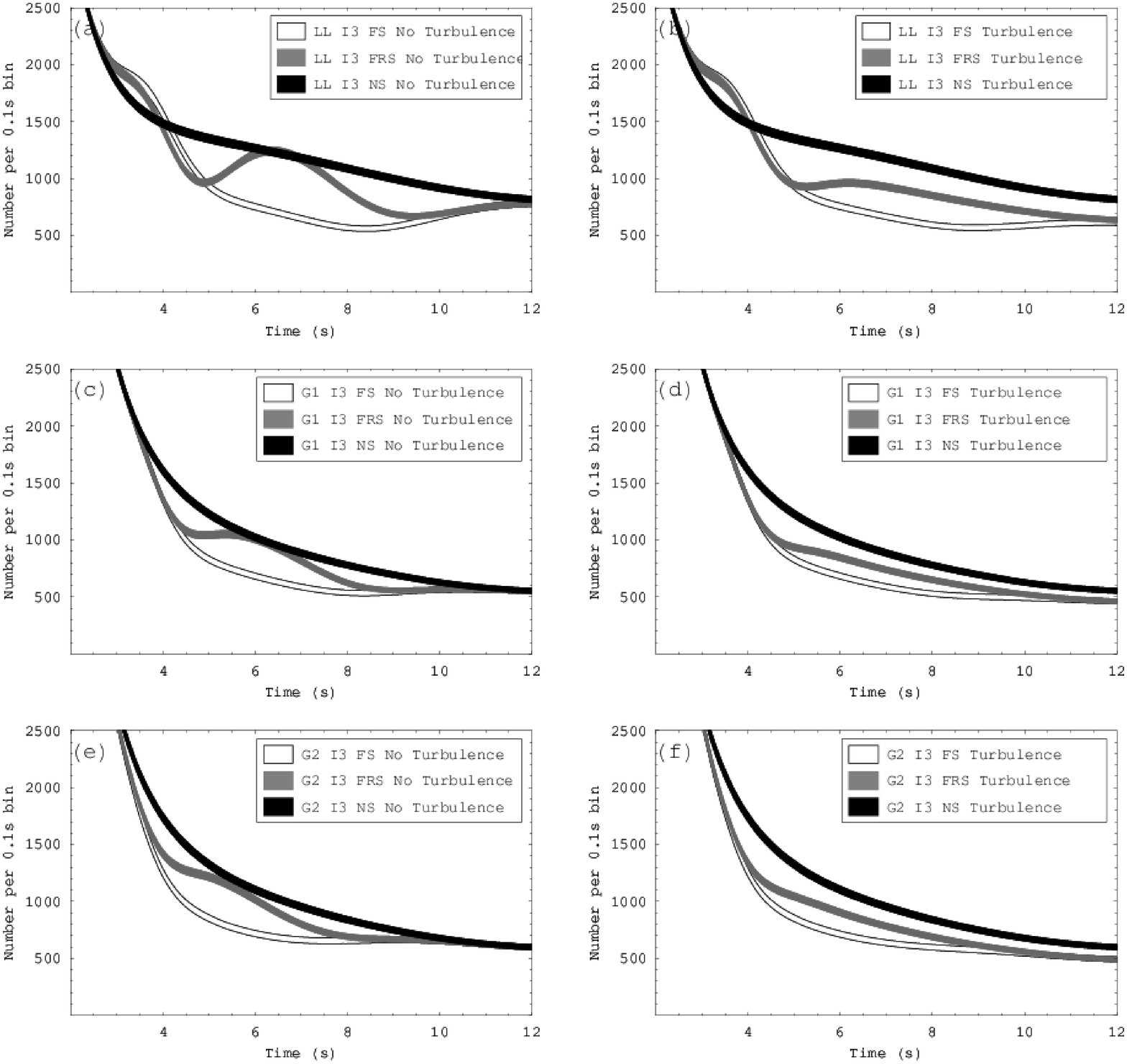}
\end{center}
\caption{The number of events in 100ms bins, for the cases of a
forward shock (FS), a forward and reverse shock (FRS) and no shock
(NS): (a)LL no turbulence, (b)LL with Turbulence, (c)G1 no
turbulence, (d) G1 with turbulence, (e)G2 no
turbulence, (f) G2 with turbulence.}\label{fig:3number}
\end{figure}

We first consider the case of a \textquotedblleft
standard\textquotedblright\ supernova at 10 kpc from earth.
If the mass hierarchy of the neutrinos is inverted, the flux of
anti-electron neutrinos in the detector is given by
\begin{equation}
F_{e}=F_{x}^{0}+\bar P_{ee}(F_{e}^{0}-F_{x}^{0}) \label{eq:flux1}
\end{equation}%
where $\bar P_{ee}$ is given by Eq. (\ref{eq:3nupee}).
%
For the values of $\sin^2\theta_{13}\gtap 10^{-3}$ that we assume
throughout this paper, $P_{13} \simeq 0$
in the absence of shock and the flux of
$\bar{\nu}_{e}$ at the detector $F_{e}\simeq F^{0}_{x}$. As the
shock crosses the resonance density, $P_{13}\simeq 1$ as discussed in
section 2.2
and the flux of
$\bar{\nu}_{e}$ at the detector is given by
$F_{e}\simeq(1-|U_{e1}|^{2})F^{0}_{x}+|U_{e1}|^{2}F^{0}_{e}$.
If the average energy of $\bar{\nu}_{x}$ is larger than that of
$\bar{\nu}_{e}$ as obtained in the LL simulations, the
average energy will decrease as a result of the shock effect.
This can be seen from Fig. \ref{fig:3avenergy}(a), where we
show the time evolution of the average energy
of the neutrinos detected in a
megaton water \chr detector. The band shows the statistical
error expected in the measured average energy.
For no shock (NS) the average energy remains almost
constant with time.
On the other hand the average energy decreases and hence shows
``bumps'' in the time profile
as the shock crosses the position of the $\ma$ driven resonance.
We get a single bump for the forward shock only and
double bump when the reverse shock is also present.
This is due to the shape of the flip
probability shown in Figs. \ref{fig:forward} and
\ref{fig:forwardandback}.
Also shown (in Fig. \ref{fig:3avenergy}(c) and
Fig. \ref{fig:3avenergy}(e)) is what is
expected if the difference between the average energies of
the $\anue$ and $\nu_x$ created in the supernova were not
so well separated. We notice that even though the effect
of the shock is less dramatic, nonetheless it is still there and
should be observable. Particularly
note that for G2 the average energy of
$\bar{\nu}_{e}$ and $\bar{\nu}_{x}$ are equal, therefore the change
in the average energy measured by the detector
is due to the change in the number of detected
neutrinos.
In Fig. \ref{fig:3number} we show the
total number of events expected
for the LL (panel (a)), G1 (panel (c)) and G2 (panel (e))
cases, with and without
the shock effects. Since the cross section for the
detection cross section increases quadratically with energy, the
number of events will decrease when the shock crosses the
$\ma$ resonance density and this results in lowering of the
number of events and hence again
gives single bump and double bump
for the forward and forward+reverse shock
cases respectively, as a function of time.

The corresponding
results taking into account the effects of turbulence are shown
in right-hand panels of
Figs. \ref{fig:3avenergy} and \ref{fig:3number}.
Panels (b), (d) and (f)
in these figures correspond to the LL, G1 and G2
simulations respectively.
The inclusion of turbulence changes the
signal for $t\gtrsim5$s. This is when the resonant density is in the
turbulent region.
As discussed before,
the system is largely depolarized giving $P_{13}\simeq1/2$.
The flux at the detector when the shock crosses the
resonance region is now given by
$F_{e}\simeq(1-0.5|U_{e1}|^{2})F^{0}_{x}+0.5|U_{e1}|^{2}F^{0}_{e}$.
For the same arguments as before, the average energy is lower than
the case of no shock, however the effect now is much less than the case
where turbulence was not considered.
Therefore, the single and
double bump features expected from shock effects get smeared
out to a large extent. However, we still see a non-trivial
variation of the expected average energy in the detector
as a function of time, both for the LL and
G1/G2 simulations.


\subsection{Three active and two sterile neutrinos}

\begin{table}[tbp]
\begin{center}
\begin{tabular}{|c|c|c|}
\hline Hierarchy & $F_{no shock}$ & $F_{shock}$\\[2ex] \hline
N2+N3 & $|U_{e1}|^{2}F^{0}_{e}$ & $P_{24}P_{25}|U_{e2}|^{2}F^{0}_{x}$ \\[2ex]
N2+I3 & $|U_{e3}|^{2}F^{0}_{e}+(|U_{e4}|^{2}+|U_{e5}|^{2})F^{0}_{x}$ & $%
|U_{e1}|^2 P_{13}F_{e}^{0}+((P_{24}+P_{25})|U_{e2}|^{2}$\\
& & $+P_{24}P_{25}
(|U_{e1}|^{2}-|U_{e2}|^{2}))F_{x}^{0}
$ \\[2ex]
H2+N3 & $|U_{e4}|^{2}F^{0}_{e}+(|U_{e3}|^{2}+|U_{e5}|^{2})F^{0}_{x}$ & $%
P_{24}|U_{e1}|^{2}F^{0}_{e}+P_{25}|U_{e2}|^{2}F^{0}_{x}$ \\[2ex]
H2+I3 & $|U_{e2}|^{2}F^{0}_{x}$ & $P_{25}|U_{e1}|^{2}F^{0}_{x}$ \\[2ex]
I2+N3 & $|U_{e2}|^{2}F^{0}_{x}$ & $P_{24}P_{25}|U_{e1}|^{2}F^{0}_{e}$ \\[2ex]
I2+I3 & $(|U_{e1}|^{2}+|U_{e2}|^{2})F^{0}_{x}$ &
$-P_{13}|U_{e1}|^{2}F^{0}_{x}$
\\ \hline
\end{tabular}%
\end{center}
\caption{\label{tab:jumpprobapprox}
The flux of neutrinos in the approximation that $%
|U_{e1}|^{2},|U_{e2}|^{2}>>|U_{e3}|^{2},|U_{e4}|^{2},|U_{e5}|^{2}$ and $%
P_{1i},P_{3i}\simeq P_{2i}$, where i=4 or 5. The total flux in
presence of shock is given by
$F_e = F_{noshock} + F_{shock}$, where $F_{noshock}=F_e$ without the
shock and $F_{shock}$ is the extra component due to the shock effect.}
\label{flipprobsimplified}
\end{table}

For the case where we have two extra light sterile neutrinos in addition
to the three active ones, the situation gets a lot more involved
due to the possibility of multiple resonances as discussed
in section 2.1. The effect of the shock is also richer here
since the shock passes through the different resonance densities
at different times.
In Table \ref{tab:jumpprobapprox}
the resultant neutrino flux at the detector is given in terms
of the original neutrino fluxes produced, for the different
neutrino mass spectra considered, for both with and without
shock effects. We reiterate that we have chosen mixing
angles such that in absence of shock effects all resonances
are adiabatic. The effect of the shock is to turn the
adiabatic resonant transition into non-adiabatic ones, as
discussed before.
For the case where sterile neutrinos are present,
the change in number of events
are characterized
by the oscillations of active neutrinos into sterile
species and vice-versa. Thus in this case,
difference in the
initial neutrino energy spectra is not a pre-requisite
for observing resonant oscillations and
shock effects, unlike in the
case of 3 active neutrinos only.
As the shock moves in time, its effect
is imprinted on the time dependence of the neutrino signal,
as in the case of 3 active neutrinos only.
However, with sterile neutrinos there
are further modulations in the signal because
the shock passes through the
additional sterile resonances. The number of resonances is dependent
on the number of sterile neutrinos as well as the
neutrino mass spectrum and
therefore the expected signal is often different for each scenario,
as described in detail in \cite{us}. In fact, as we will see,
even when the shock passes through the $\ma$ driven
active resonance, the time profile and energy spectrum of the
observed neutrinos in many of the possible mass spectrum
listed in Eqs. (\ref{eq:n2n3})-(\ref{eq:i2i3}),
are different than what is expected for the
case of 3 active neutrinos only.
Typically, the shock crosses the
multiple ``sterile
resonances'' at very early times ($t \ltap 2$ sec), while
it crosses the $\ma$ driven ``active
resonance'' at later times ($t \gtap 3$ sec).
Therefore in what follows, we will present results separately
at late and early times to show clearly the effect of the shock wave
on the neutrino signal in the detector. For later times, we will
consider time bins of 100 ms, while at earlier times since the
time dependence is much more sharp, we present our results for
smaller time bins of 10 ms.

\subsection{Neutrino signal at late times}

\begin{figure}[p]
\begin{center}
\includegraphics[width=17cm]{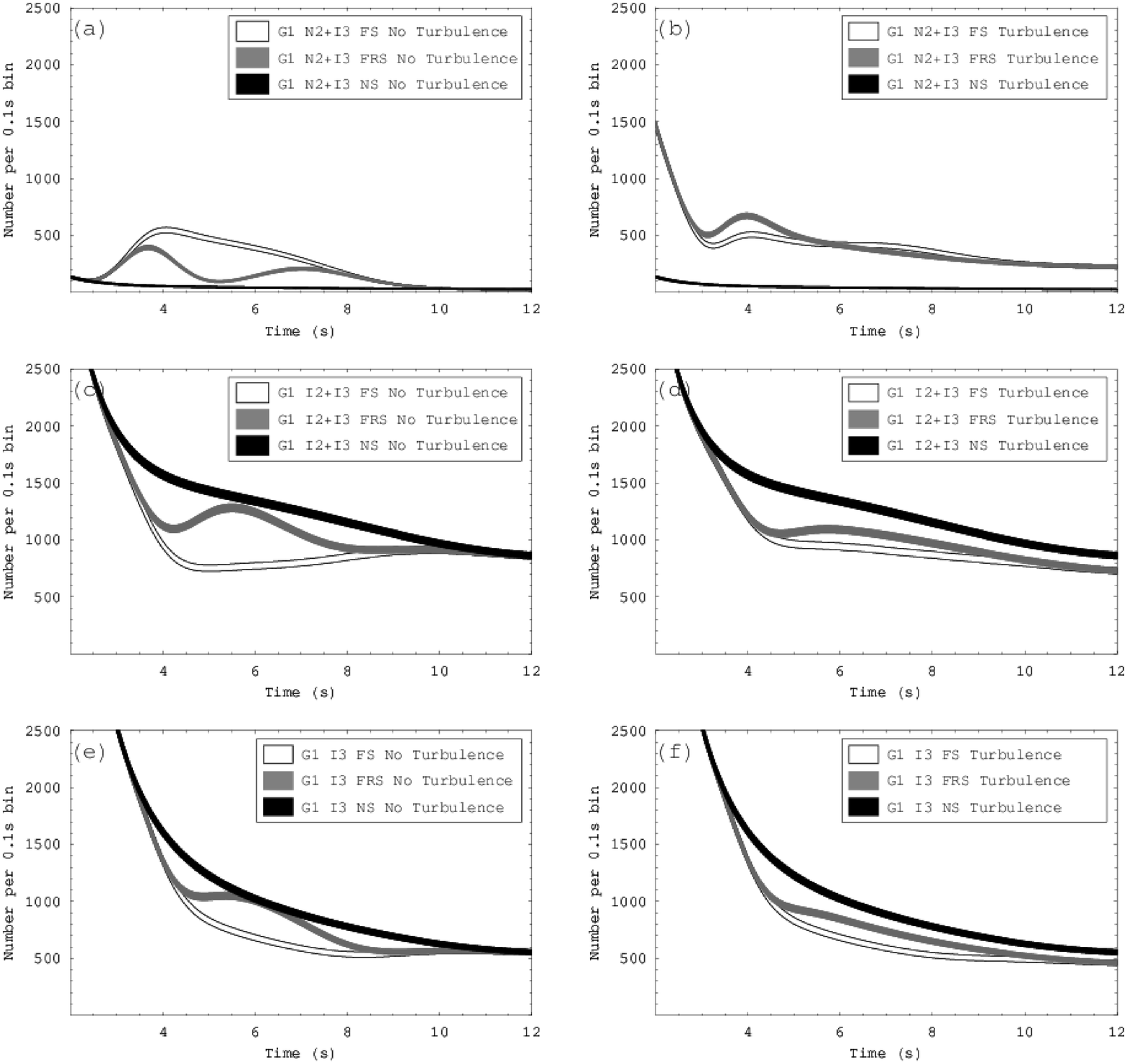}
\end{center}
\caption{The number of events in 100ms bins, for the cases of a
forward shock (FS), a forward and reverse shock (FRS) and no shock
(NS): (a)N2 + I3 no turbulence, (b)N2 + I3 with turbulence, (c)I2 +
I3 no turbulence, (d)I2 + I3 with turbulence, (e)I3 no turbulence,
(f)I3 with turbulence.}\label{fig:largenumber}
\end{figure}

\begin{figure}[p]
\begin{center}
\includegraphics[width=17cm]{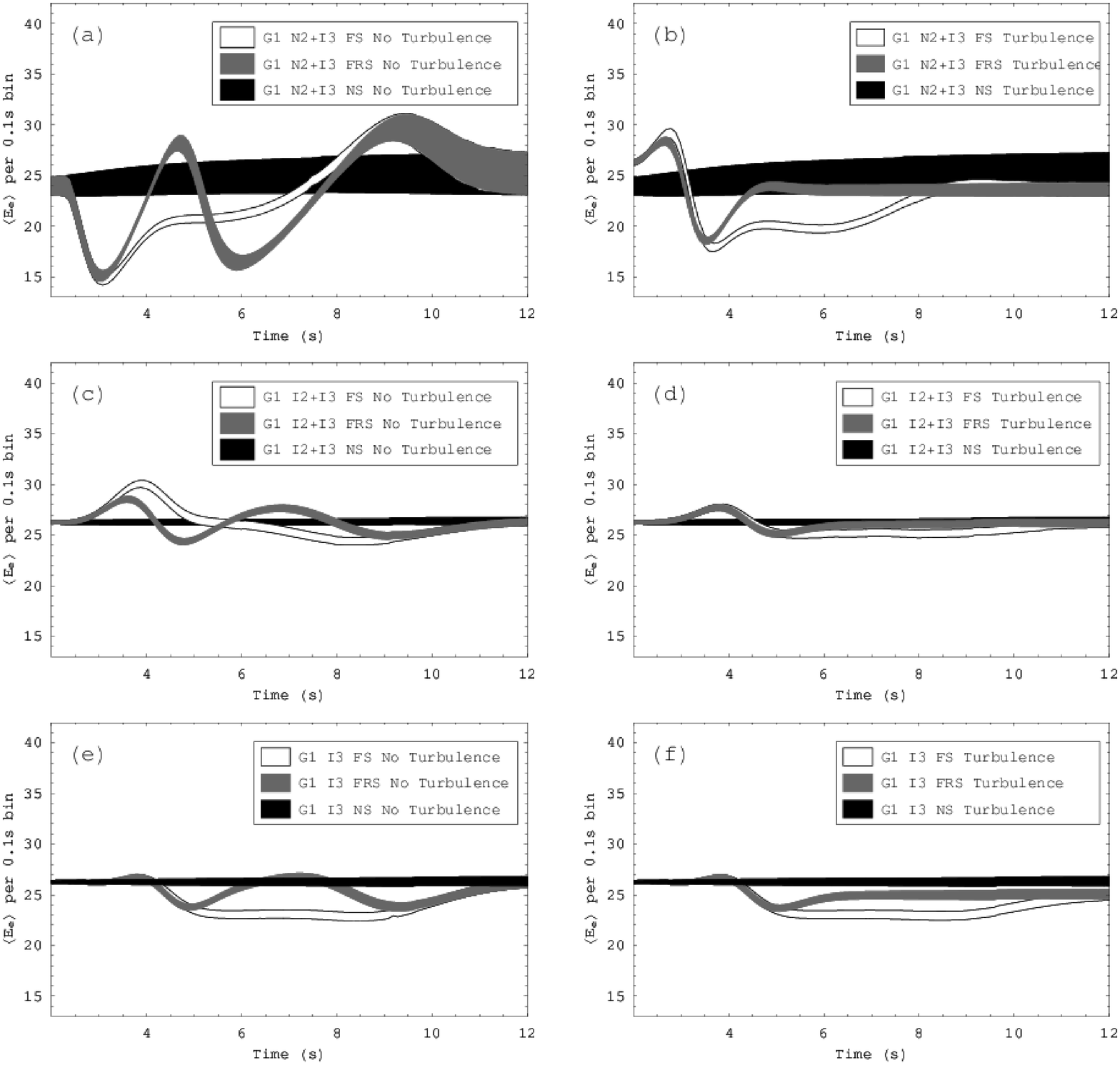}
\end{center}
\caption{The average energy in 100 ms bins, for the cases of a
forward shock (FS), a forward and reverse shock (FRS) and no shock
(NS), with the initial energy spectra G1 as discussed in the text:
 (a)N2 + I3 no turbulence, (b)N2 + I3 with turbulence, (c)I2 + I3
no turbulence, (d)I2 + I3 with turbulence, (e)I3 no turbulence,
(f)I3 with turbulence.}\label{fig:largeavenergy}
\end{figure}

We begin by discussing the evolution of the
expected neutrino event rate
and average energy at later
times. The total number of events for the different
mass spectra are
shown in Fig. \ref{fig:largenumber} while the expected average
energies are shown in Fig. \ref{fig:largeavenergy}, at
times between $t= 3-12$s.
These plots show the impact
on the signal when the shock wave passes through the $\ma$
resonance between active neutrinos.
In panels (a)-(d) of Figs. \ref{fig:largenumber}
and \ref{fig:largeavenergy}
we show only the mass spectrum cases where $\ma <0$,
in which we have resonance and hence also shock effects.
Which mass spectra will get the shock effects can be easily
seen from Table \ref{tab:jumpprobapprox}. Since at late times
the resonance that gets affected by the shock is the
$\ma$ driven resonance between the 1-3 states, the relevant
jump probability involved is $P_{13}$. All other jump
probabilities are zero here, from our choice of the mixing angles.
We can see from  Table \ref{tab:jumpprobapprox} that the
N2+I3 and I2+I3 are the only cases which will get affected.
For all the other mass spectra possible,
we do not expect any modulation in the signal at late times
due to shock effects. Hence we show only the signal for
the N2+I3 and I2+I3 cases.
For comparison we show in the
last 2 panels of these figures the case of only
active neutrinos (I3).

In the approximation that
$|U_{e3}|^2$, $|U_{e4}|^2$ and $|U_{e5}|^2$ can be neglected
in comparison to $|U_{e1}|^2$ and $|U_{e2}|^2$, we note
from Table \ref{tab:jumpprobapprox}
that in absence of shock,
\be
I3 &\Rightarrow& {F_e} \simeq F_x^0~,\label{eq:fluxNSI3}\\
N2+I3 &\Rightarrow& {F_e} \simeq 0~,\label{eq:fluxNSN2I3}\\
I2+I3 &\Rightarrow& {F_e} \simeq F_x^0\label{eq:fluxNSI2I3}~,
\ee
whereas when the shock passes through the 1-3 resonance the fluxes
are modified to,
\be
I3 &\Rightarrow& {F_e} \simeq
F_x^0(1-P_{13}|U_{e1}|^2)+P_{13}|U_{e1}|^2F_e^0~,\label{eq:fluxSI3}\\
N2+I3 &\Rightarrow& F_e \simeq |U_{e1}|^2P_{13}F_e^0~,\label{eq:fluxSN2I3}\\
I2+I3 &\Rightarrow& {F_e} \simeq (1-P_{13}|U_{e1}|^2)F_x^0
\label{eq:fluxSI2I3}~.
\ee
By comparing Eqs. (\ref{eq:fluxNSI3})-(\ref{eq:fluxNSI2I3})
we note that in absence of shock,
while the fluxes are same for I3 and I2+I3 spectra,
N2+I3 predicts almost zero fluxes.
The black bands in Fig. \ref{fig:largenumber} corroborate
the above statement.
When the shock passes through the $\ma$ resonance it increases
the $\anue$ flux for N2+I3, while it decreases the same for
I2+I3, as can be seen by comparing Eqs. (\ref{eq:fluxNSN2I3})
and (\ref{eq:fluxNSI2I3}) with Eqs. (\ref{eq:fluxSN2I3})
and (\ref{eq:fluxSI2I3}). In the case of I3, the resultant
flux is an admixture of a fraction of the initial $\anue$ and
$\nu_x$ fluxes. However, since the average energy of initial $\anue$
flux was smaller, the total number of events goes down in the
detector when the shock passes through the 1-3 resonance even
in this case. We stress that the sterile cases are qualitatively
different from the I3 case, since for them the net number flux sees
a big increase or decrease due to shock.  Therefore, while
for I3, the effect of the shock wave comes predominantly through the
difference in the average energy of $\anue$ and $\nu_x$,
for sterile cases we see a combined effect coming from a direct
change in the number of neutrinos arriving on earth as well as
the difference in the energy spectra of the different species.

For all the mass spectra shown, under the
assumption that there is no turbulence,
typically a
single bump is observed for a forward shock and a double bump for a
forward and reverse shock, as expected from the shape of the flip
probability shown in Figs. \ref{fig:forward} and
\ref{fig:forwardandback}.
The effect of taking the turbulence into account is to smear these
sharp changes in the oscillation probability due to the
effective depolarizing of the resonances.
With turbulence, a single bump
is observed, followed by a region in which the number of events are
typically different from what is expected for no turbulence. This
can be seen in the right-hand panels in Fig. \ref{fig:largenumber}.

The expected average energies for the N2+I3 and I2+I3 cases
also show a very striking evolution with time at $t\gtap 3$s,
which is very different from that predicted by I3.
In the case of I3, the average energy decreases in
presence of shock as expected, since the shock reduces the
conversion probability of the high energy $\nu_x$ into $\anue$,
thereby decreasing the average energy.
Thus it usually shows 1 sharp
decrease for forward shock and 2 sharp falls for forward+reverse
shock.
On the
other hand both N2+I3 and I2+I3 predict that the average
energy fluctuates on both sides of the average energy expected
in absence of shock. The key issue to note here is that the
position of resonance is determined by the energy of the neutrino.
The higher (lower) energy neutrinos go through the resonance
at lower (higher) density. Since the shock moves from higher to
lower densities in time, the lower energy neutrinos are affected
by the shock earlier than the higher energy ones.
For the sterile cases, the effect is more subtle. Here the
time evolution of the average energy is a combined
effect of the change in the total flux as well as the
energy dependence of the resonance position.
For the I2+I3 case
since resonance happens for lowest energy neutrinos
first, the effect
of the shock is to reduce
them in the flux (cf. Eq.(\ref{eq:fluxNSI2I3}) and
(\ref{eq:fluxSI2I3})), thereby increasing the average energy.
Eventually, the shock goes through higher
energy resonances, and this then reduces the average energy.
For the N2+I3 case also
the resonance happens for the lowest energy neutrinos
first, but now the shock effect increases them in the flux and
thereby decreases the average energy. Eventually,
the shock goes through the high energy resonances,
increasing the average energy.

The right-hand panels of Fig. \ref{fig:largeavenergy} show the average
energy after turbulence is taken into consideration.
The presence of turbulence has a typical effect on the
time evolution of both the total number of events and the
average energy deposited by the (anti)neutrinos. One can see
that without turbulence both the average energy and number of events
after the passage of (both) shock(s) go back to its
pre-shock value. This is because the $\bar P_{ee}$ before and
after the shock are exactly the same in this case.
However, since the turbulence changes the flip probability
permanently to $P_{ij}$ behind the shock, the
number of events and average energy are consistently lower
once the shock crosses the resonance point.


\subsection{Neutrino signal at early times}

\begin{figure}[p]
\begin{center}
\includegraphics[width=17cm]{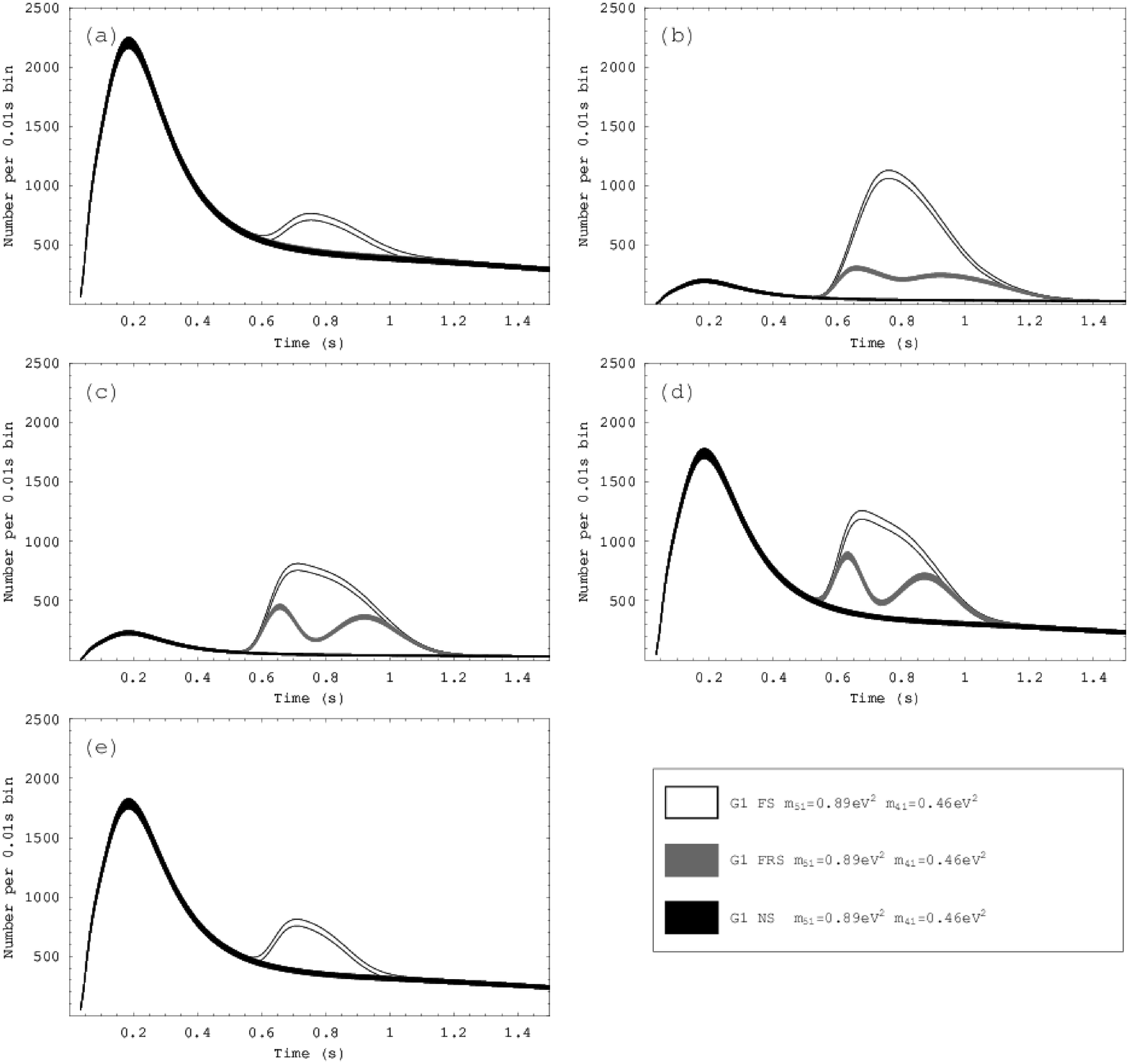}
\end{center}
\caption{The number of events in 10 ms bins, for the cases of
$\Delta m^{2}_{41}=0.46MeV^{2}$ and $\Delta m^{2}_{51}=0.54MeV^{2}$,
with no turbulence, a forward shock (FS), a forward and reverse
shock (FRS) and no shock (NS), with the initial energy spectra G1 as
discussed in the text:(a)N2 + N3, (b)N2 + I3, (c)H2 + N3, (d)H2 +
I3, (e)I2 + N3.}\label{fig:smallnumber}
\end{figure}

\begin{figure}[p]
\begin{center}
\includegraphics[width=17cm]{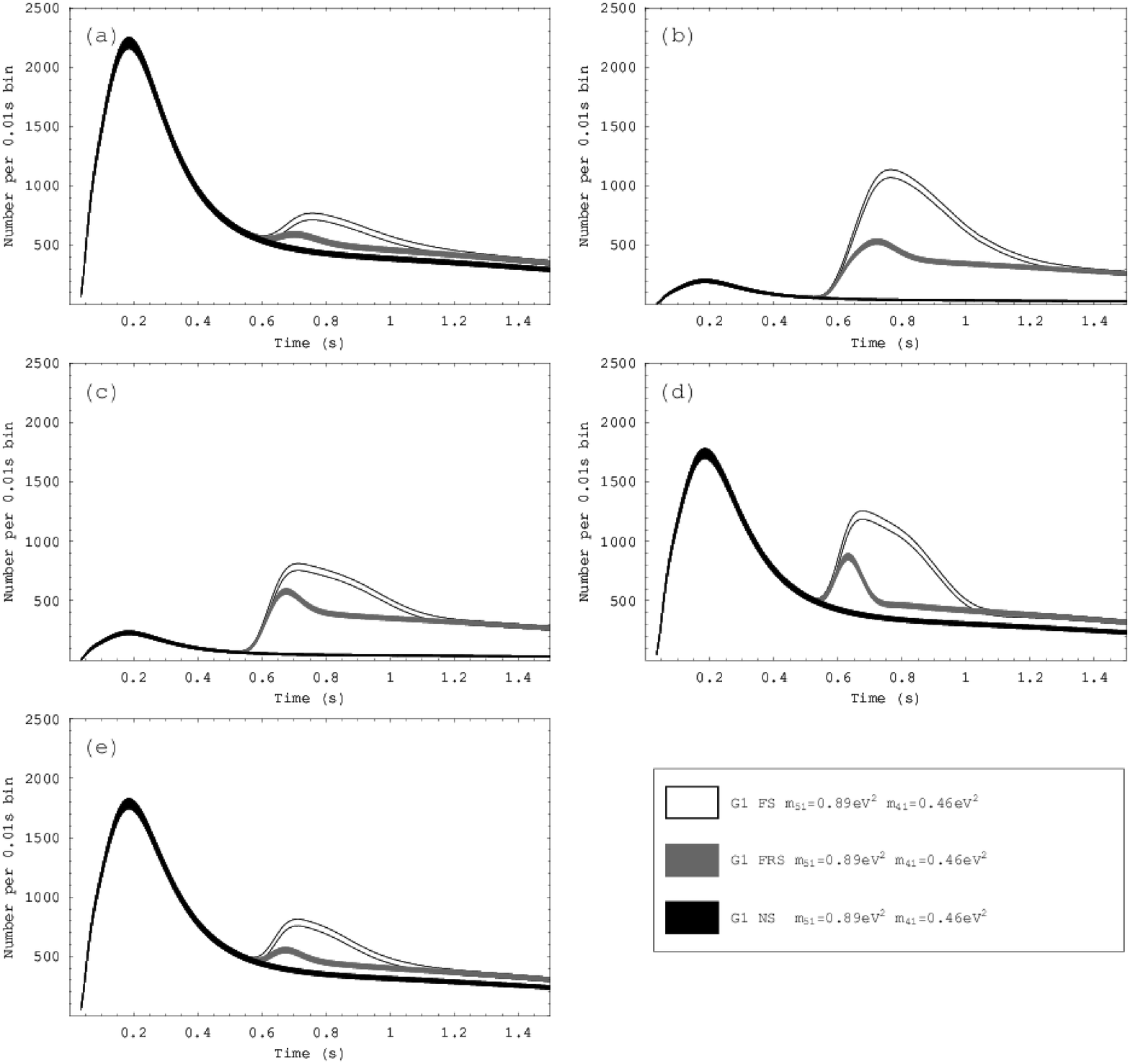}
\end{center}
\caption{The number of events in 10 ms bins, for the cases of
$\Delta m^{2}_{41}=0.46MeV^{2}$ and $\Delta m^{2}_{51}=0.54MeV^{2}$,
with turbulence, a forward shock (FS), a forward and reverse shock
(FRS) and no shock (NS), with the initial energy spectra G1 as
discussed in the text:(a)N2 + N3, (b)N2 + I3, (c)H2 + N3, (d)H2 +
I3, (e)I2 + N3.}\label{fig:smallnumberturb}
\end{figure}

\begin{figure}[p]
\begin{center}
\includegraphics[width=17cm]{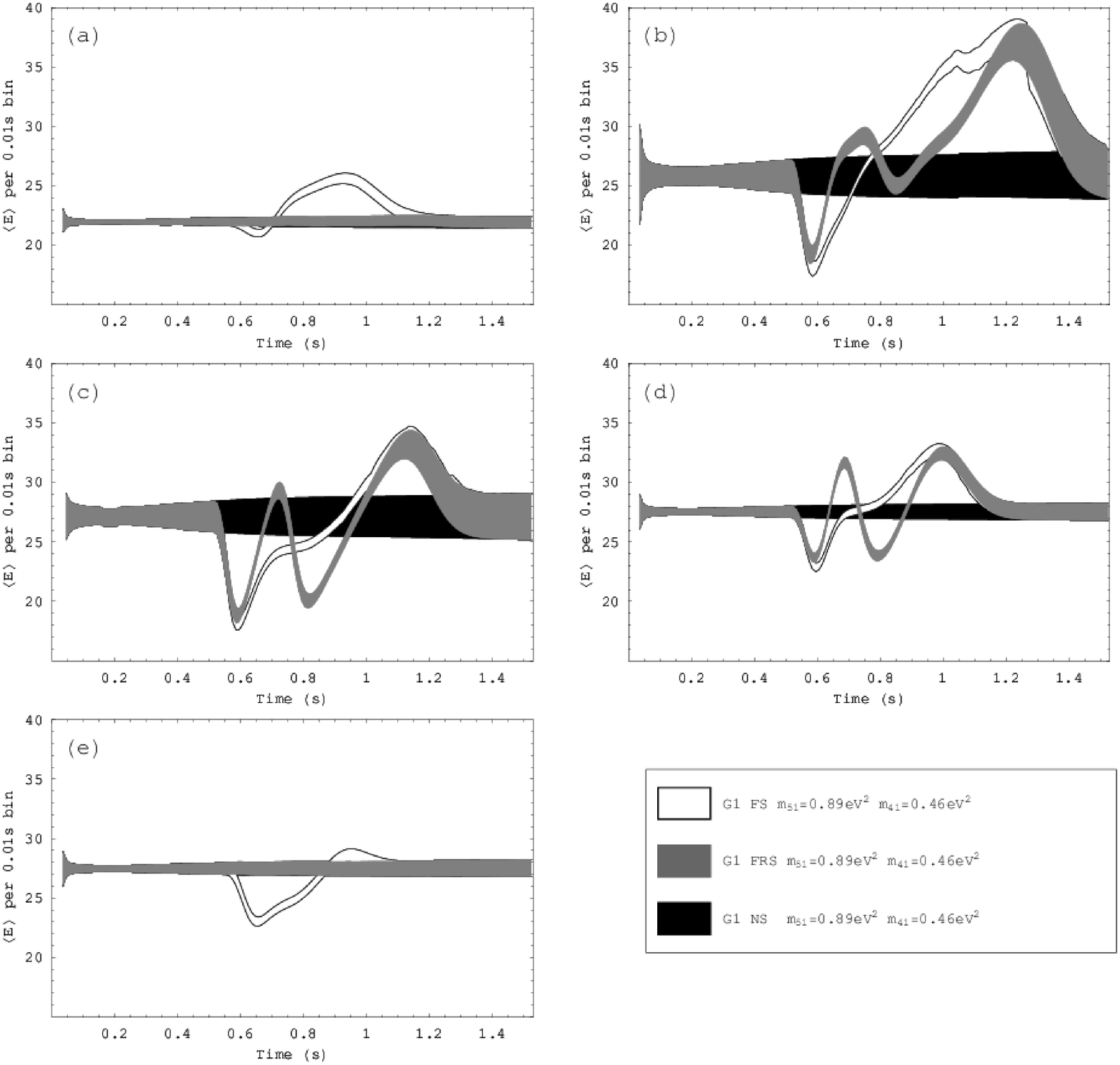}
\end{center}
\caption{The average energy in 10 ms bins, for the cases of $\Delta
m^{2}_{41}=0.46MeV^{2}$ and $\Delta m^{2}_{51}=0.54MeV^{2}$, with no
turbulence, a forward shock (FS), a forward and reverse shock (FRS)
and no shock (NS), with the initial energy spectra G1 as discussed
in the text:(a)N2 + N3, (b)N2 + I3, (c)H2 + N3, (d)H2 + I3, (e)I2 +
N3.}\label{fig:smallavenergy}
\end{figure}

\begin{figure}[p]
\begin{center}
\includegraphics[width=17cm]{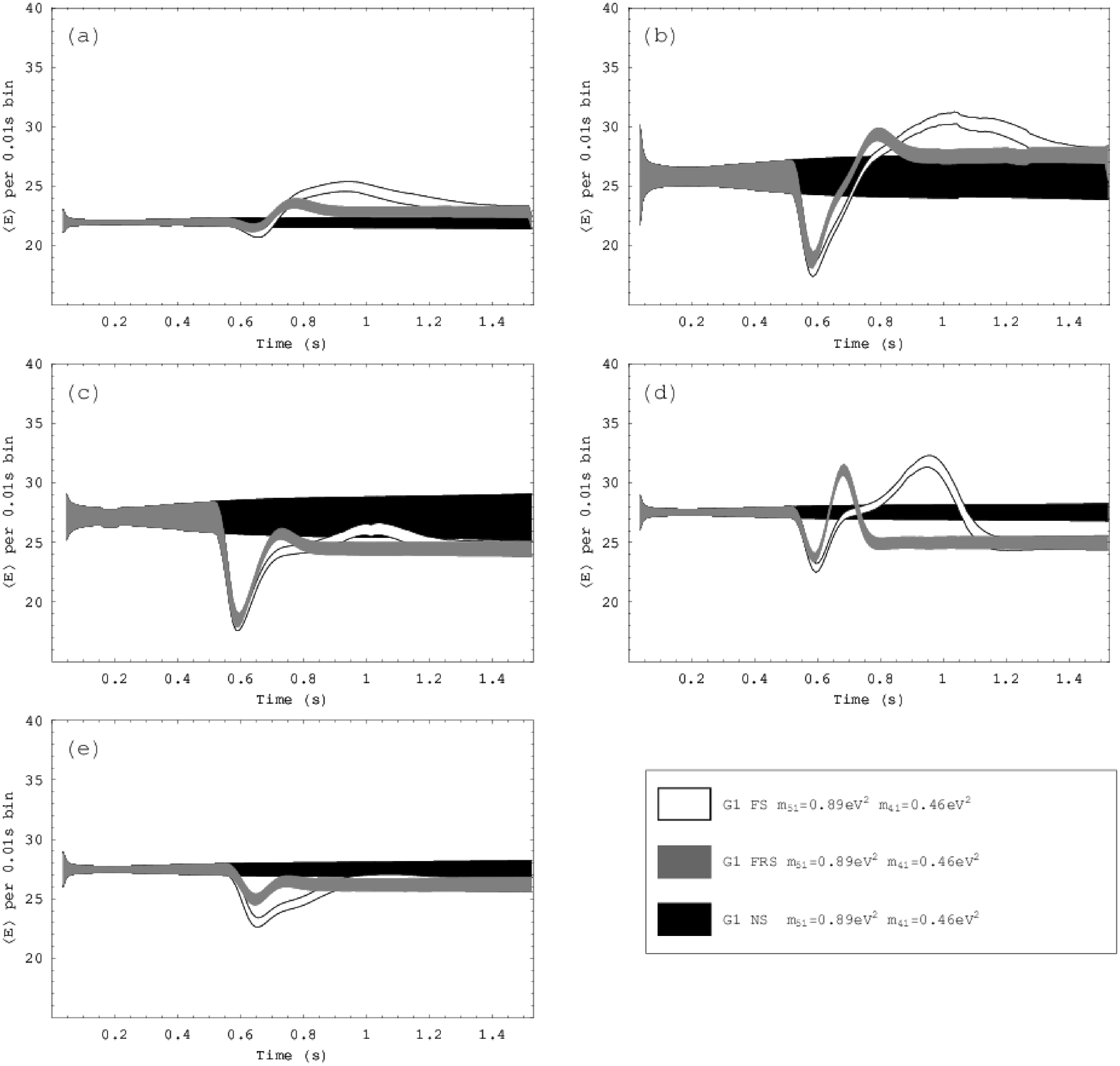}
\end{center}
\caption{The average energy in 10 ms bins, for the cases of $\Delta
m^{2}_{41}=0.46MeV^{2}$ and $\Delta m^{2}_{51}=0.54MeV^{2}$, with
turbulence, a forward shock (FS), a forward and reverse shock (FRS)
and no shock (NS), with the initial energy spectra G1 as discussed
in the text:(a)N2 + N3, (b)N2 + I3, (c)H2 + N3, (d)H2 + I3, (e)I2 +
N3.}\label{fig:smallavenergyturb}
\end{figure}

In Figs. \ref{fig:smallnumber} and  \ref{fig:smallavenergy}
we show the number of predicted events and average energies
as a function of time (in absence of turbulence)
for the 5 different mass spectra,
between $t=0.1-1.5$s, in short time bins of 10 ms.
Figs. \ref{fig:smallnumberturb} and \ref{fig:smallavenergyturb}
show the corresponding results when turbulence is taken into
account. We do not show results for the I2+I3 spectra here since,
as can be seen from Table \ref{tab:jumpprobapprox},
it depends only on $P_{13}$ which will have shock effects
only at the $\ma$ resonance at late times.

As in the previous subsection,
Fig. \ref{fig:smallnumber} can be understood in terms of
the flux predictions in the presence and absence of the shock, given
in Table \ref{tab:jumpprobapprox}. The only difference is
that at early times the shock passes through the sterile
resonances and hence we expect contributions coming from the
jump probabilities associated with the sterile resonances.
In particular, we note that in the absence of shock, the
predicted flux on earth is almost zero for N2+I3 and H2+N3,
while for H2+I3 and I2+N3 we expect the flux to be
$|U_{e2}|^2F_x^0$ and for N2+N3 it is predicted to be
$|U_{e1}|^2F_x^0$. Since $|U_{e1}|^2>|U_{e2}|^2$, the
expected signal is larger for N2+N3. These features are
evident from Fig. \ref{fig:smallnumber}. As a result of the
shock we see a modulation in the resultant signal
visible as bumps in the figure. Note that in all
the 5 cases shown, the shock effect increases the number of
events.
The shock powered modulations get affected when turbulence is
taken into account and the corresponding results
are shown in Fig. \ref{fig:smallnumberturb}.

Fig. \ref{fig:smallavenergy} shows how the average energy
evolves at early times. As we had seen at later times,
the effect of the shock wave is to make the
average energy fluctuate on both sides
of the corresponding
static density case. The reason for the initial average
energy decrease and subsequent increase is also the same as
that discussed in the previous subsection. Fig.
\ref{fig:smallavenergyturb} shows the average
energy evolution when the effect of turbulence is
considered. The rapid fluctuations are mellowed due to the presence of turbulence, however we
still see statistically significant fluctuations in
the average energy at very early times, a feature
which, if observed, would provide an almost
model independent signal of the presence of
extra sterile neutrino species which are mixed mildly
with the active neutrinos.

\begin{figure}[p]
\begin{center}
\includegraphics[width=17cm]{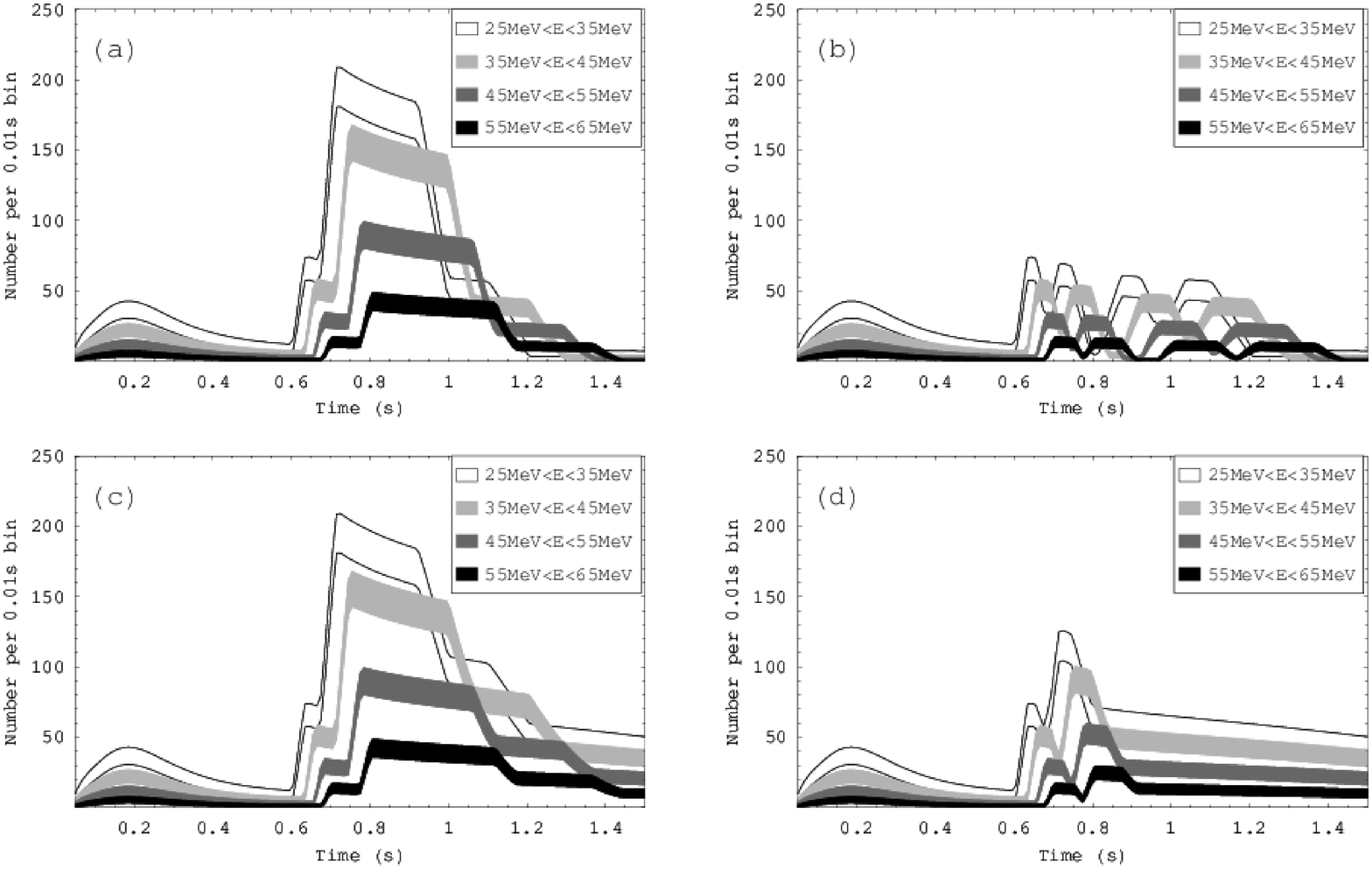}
\end{center}
\caption{The number of events in 10 ms and 10 MeV bins, for N2 +I3
and the cases of a forward shock (FS), a forward and reverse shock
(FRS), with the initial energy spectra LL as discussed in the
text:(a)FS no turbulence, (b)FRS no turbulence, (c)FS with
turbulence, (d)FRS with turbulence.}\label{fig:ebin}
\end{figure}

As the water \chr
detector can measure the energy of the incoming neutrino
rather efficiently, the
events can be binned in energy as well as time.
In Fig.  \ref{fig:ebin} we present the number of
events expected in energy bins of width 10 MeV and
time bins of 10 ms.
We show results for the N2+I3 spectrum only as an
example. This figure is shown
for only early times to illustrate the effect of the
sterile resonances.
As the resonance densities are energy
dependent, the propagation of the shock wave can be observed when the
events are binned in energy and time. The shock wave crosses the
resonant density corresponding to neutrinos with lower energies
first, therefore the characteristic 'bumps' are observed in the lower
energy bins first. The presence of such energy and time dependent
bumps in the resultant signal in the detector provides a
'smoking gun' signal for sterile neutrinos.

\section{Comparison and Discussions}

\begin{table}[h]
\begin{center}
\begin{tabular}{|c|c|c|c|c|}
\hline
Spectrum & \multicolumn{2}{|c|}{{\rule[0mm]{0mm}{6mm}Early Times}}
& \multicolumn{2}{|c|}{{\rule[0mm]{0mm}{6mm}Late Times}} \cr \cline{2-5}
& Events & Shock Effect & Events & Shock Effects \cr \hline\hline
N3 & L & ${\times}$ & L &  ${\times}$\cr
I3 & L & ${\times}$ & L &  ${\checkmark}$\cr
N2+N3 & L & ${\checkmark}$ & L &  ${\times}$\cr
N2+I3 & S & ${\checkmark}$ & S &  ${\checkmark}$\cr
H2+N3 & S & ${\checkmark}$ & S &  ${\times}$\cr
H2+I3 & L & ${\checkmark}$ & L &  ${\times}$\cr
I2+N3 & L & ${\checkmark}$ & L &  ${\times}$\cr
I2+I3 & L & ${\times}$ & L &  ${\checkmark}$\cr
\hline
\end{tabular}%
\end{center}
\caption{\label{tab:signal}
The expected trends in the neutrino signal in a megaton
water \chr detector at early and late times. The symbols
L and S signify large and small number of events respectively.
The presence or absence of shock effects are shown by
the ${\checkmark}$ and ${\times}$ symbols.
}
\end{table}

In Table \ref{tab:signal} we show the
model independent characteristic features in the expected
signal for the eight different cases considered in this paper, the two only active cases (N3 and I3) and the six active plus sterile cases.
The time interval is divided into early ($t \ltap 2$ s) and
late ($t\gtap 3$ s) times, as before. For each case we
state if we expect large (L) or very small (S) number of
events in the 2 time zones. We also indicate in the table
whether we expect shock induced sharp modulations in the
signal or not for each of the possible mass spectra.
By comparing the predictions for all the cases, we can infer
the following:
\begin{itemize}

\item \begin{enumerate} \item If large events are seen at early times,
\item shock effects are not seen at early times,
\item a large number of events are seen at late times and
\item shock effects are not seen at late times \end{enumerate} then
we {\it must} have the N3 spectrum.

\item \begin{enumerate} \item If very few events are seen at early times,
\item  shock effects are seen at early times,
 \item very few events are seen at late times and
\item  shock effects are seen at late times\end{enumerate}  then
we {\it must} have the N2+I3 spectrum.

\item \begin{enumerate} \item  If very few events are seen at early times,
\item shock effects are seen at early times,
\item  very few events are seen at late times and
\item shock effects are not seen at late times \end{enumerate} then
we {\it must} have the H2+N3 spectrum.

\item  \begin{enumerate} \item If a large number of events are seen at early times,
\item  shock effects are seen at early times,
\item  a large number of events are seen at late times and
\item  shock effects are not seen at late times \end{enumerate}  then
we {\it could} have either N2+N3 or H2+I3 or I2+N3 spectrum.

\item  \begin{enumerate} \item If a large number of events are seen at early times,
\item  shock effects are not seen at early times,
\item a large number of events are seen at late times and
\item shock effects are seen at late times \end{enumerate}  then
we {\it could} have either I3 or I2+I3 spectrum. This degeneracy can be split by measurng the energy dependence
(c.f. Figure 6 (c) and (d)).

\end{itemize}

Therefore, from the Table \ref{tab:signal} we conclude that,
irrespective of model uncertainties, the
presence of sterile neutrinos can be easily proved from modulation
of the signal due to shock effects at early time. Only the
I2+I3 case does not give any time modulation, even though the
sterile neutrinos are present in this case. In addition, from
the point-wise discussion above we can see how well we can
distinguish one mass spectrum from the other. N3, N2+I3 and
H2+N3 can be uniquely determined by comparing early time behavior
of the signal with its late time behavior. The N2+N3,
H2+I3 and I2+N3 can be separated from the rest, but since they
predict similar trends at early and late times, one will be need
a more careful model dependent study to unambiguously
disentangle them from each other. Similarly, though I3 and I2+I3
can be separated from the other spectra, one will need
a more careful analysis to distinguish between the two.

\section{Summary and Conclusions}

Model independent information about the neutrino mass spectrum
can be obtained through observation of signatures of
supernova shock wave(s) on the resultant neutrino signal in
terrestrial detectors. In particular, such signals probe the existence of
extra sterile neutrino through their possible resonant transition
with active neutrinos inside the supernova. Oscillations
between sterile neutrino and active ones are characterized by
unique signatures in the final neutrino energy spectrum
as well as their evolution with time. While the
impact of the presence of sterile neutrinos on the
time evolution of the neutrino is due to shock effects,
their impact on the resultant neutrino flux and spectra
comes from neutrino oscillations both with and without shock
effects.

Concerns had been raised recently over the
observability of the shock effects due to the
turbulent density variations following the shock wave which prove to be very important.
In this paper we made a detailed study of the
turbulent shock effects
on the neutrino induced galactic supernova signal expected in
megaton water \chr detector. Water detectors can give
information on the number,
the energy spectrum, as
well the time profile of the arriving (anti)neutrinos.
We have shown that the impact of the
supernova shock waves
is evident in the neutrino signature
in megaton water \chr detectors in all of these,
making this class of detectors extremely good
for studying shock effects. We have considered the
impact of the turbulence left behind by the shock wave and
have seen that although the shock effect
is diluted, it is still significant and we
still expect to observe them in the megaton class of
detectors.
We have concentrated
on the discernable signatures of sterile neutrinos, which might be
mixed mildly with the 3 active neutrinos. We have illustrated the effects using sterile neutrino parameters which were chosen in a fit to all neutrino data including LSND. However we showed that the observable signals persist for much smaller mixing angles than are observable by the LSND (or MiniBOONE) experiments.  Hence our results are
relevant whatever the outcome of the MiniBOONE data.

The most striking evidence for sterile neutrinos in the
supernova
neutrino signal are sharp bumps
at $t \ltap 1$secin the observed number
flux as well as the average energy of the $\anue$ detected
through their capture on protons in megaton water detectors.
These can be caused only by sterile neutrino resonances
inside the supernova. Only the I2+I3 case for the
mass spectrum does not predict these early time
shock induced modulations
in the signal.
In addition, a model independent comparison of the
signal trend between early and late times can give us
a rather unambiguous signature on 3 of the 8 possible
mass spectra considered in this paper. The N3, N2+I3 and
H2+N3 cases predict unique combination of behavior at early
and late times and hence can be determined model
independently from the observations. The remaining 5 cases
can be classed into 2 categories depending on their
combination of predicted trends at early and late times.
Distinguishing the I3 from the I2+I3 spectra would require
a more careful model dependent analysis of the future supernova
neutrino data. Similarly, the N2+N3, H2+I3 and I2+N3 can be
separated from the relative differences in their predictions of
average energy and number of events as a function of time.

One important byproduct of the turbulent effects involving active-sterile neutrino mixing is that, even if the energy spectrum and luminosisites of the active neutrinos are initially the same, the depolarising effect of the turbulence for the active sterile resonances in the wake of the shock front will make the active neutrino spectra spectra and luminosities significantly different. As a result the atmospheric neutrino resonant effects involving the active neutrinos in the presence of sterile neutrinos may be expected to give rise to more significant effects in the supernova neutrino signal than is the case without sterile neutrinos.

\vskip 1cm
\noindent
{\bf Acknowledgment}\\
{We thank T. Kajita for discussion on water \chr detectors.
SC wishes to thank PPARC and the University
of Oxford for financial support during this work.}


\end{document}